\documentclass[sigconf]{aamas} 
\usepackage{balance} % for balancing columns on the final page

\usepackage{appendix}
\usepackage{bm} 
\usepackage{amsmath}
\usepackage{amsthm}
\usepackage{multirow}
\usepackage{graphicx}
\usepackage{comment}
\usepackage{subcaption}
\usepackage{subcaption}
\usepackage{graphics}
\usepackage{soul}
\usepackage[linesnumbered,ruled,vlined]{algorithm2e}
\usepackage{diagbox} 
\usepackage{multirow}
\usepackage{booktabs}

\providecommand{\jiang}[1]{\textcolor{green}{#1}}

\newcommand{\re}[1]{{\color{black}#1}}
\newcommand{\revise}[1]{{\color{black}#1}}

\newtheorem{definition}{Definition}[section]
\newtheorem{problem}{Problem Statement}[section]
\setcopyright{ifaamas}
\acmConference[AAMAS '24]{AAMAS 2024}
\copyrightyear{2024}
\acmDOI{}
\acmPrice{}
\acmISBN{}
\acmSubmissionID{242}

\begin{document}

\title[DuaLight]{DuaLight: Enhancing Traffic Signal Control by Leveraging Scenario-Specific and Scenario-Shared Knowledge}

%%% Provide names, affiliations, and email addresses for all authors.

\author{Jiaming Lu$^{\dagger}$$^{\ddagger}$}
\affiliation{
  \institution{SenseTime Research \\ ISTBI of Fudan University}
  \city{Shanghai}
  \country{China}}
\email{lujia_ming@126.com}

\author{Jingqing Ruan$^{\dagger}$$^{\ddagger}$}
\affiliation{
  \institution{SenseTime Research}
  \city{Beijing}
  \country{China}}
\email{ruanjingqing@sensetime.com}

\author{Haoyuan	Jiang$^{\S}$}
\affiliation{
  \institution{SenseTime Research \\ Baidu Inc.}
  \city{Shenzhen}
  \country{China}}
\email{jianghaoyuan@zju.edu.cn}

\author{Ziyue Li$^{\ast}$}
\affiliation{
  \institution{Germany University of Cologne \\ EWI gGmbH}
  \city{Cologne}
  \country{Germany}}
\email{zlibn@wiso.uni-koeln.de}

\author{Hangyu Mao}
\affiliation{
  \institution{SenseTime Research}
  \city{Beijing}
  \country{China}}
\email{maohangyu@senstime.com}
		
\author{Rui	Zhao}
\affiliation{
  \institution{SenseTime Research \\ Qing Yuan Research Institute of Shanghai Jiao Tong University}
  \city{Beijing}
  \country{China}}
\email{zhaorui@sensetime.com}

\thanks{$\dagger$ These authors contribute equally to this work.}
\thanks{$\ddagger$ These authors work as research interns at SenseTime Research.}
\thanks{${\S}$ Work done at SenseTime Research, now in Baidu Inc.}
\thanks{$\ast$ The corresponding author.}

%%% Use this environment to specify a short abstract for your paper.

\begin{abstract}
Reinforcement learning has been revolutionizing the traditional traffic signal control task, showing promising power to relieve congestion and improve efficiency. However, the existing methods lack effective learning mechanisms capable of absorbing dynamic information inherent to a specific scenario and universally applicable dynamic information across various scenarios. Moreover, within each specific scenario, they fail to fully capture the essential empirical experiences about how to coordinate between neighboring and target intersections, leading to sub-optimal system-wide outcomes.

Viewing these issues, we propose DuaLight, which aims to leverage both the experiential information within a single scenario and the generalizable information across various scenarios for enhanced decision-making. Specifically, DuaLight introduces a scenario-specific experiential weight module with two learnable parts: Intersection-wise and Feature-wise, guiding how to adaptively utilize neighbors and input features for each scenario, thus providing a more fine-grained understanding of different intersections. Furthermore, we implement a scenario-shared Co-Train module to facilitate the learning of generalizable dynamics information across different scenarios. Empirical results on both real-world and synthetic scenarios show DuaLight achieves competitive performance across various metrics, offering a promising solution to alleviate traffic congestion, with 3-7\% improvements. The code is available under 
\url{https://github.com/lujiaming-12138/DuaLight}.
% \url{https://anonymous.4open.science/r/DuaLight-1648}. 

\end{abstract}

%%% The code below was generated by the tool at http://dl.acm.org/ccs.cfm.
%%% Please replace this example with code appropriate for your own paper.

%%% Use this command to specify a few keywords describing your work.
%%% Keywords should be separated by commas.


%%%%%%%%%%%%%%%%%%%%%%%%%%%%%%%%%%%%%%%%%%%%%%%%%%%%%%%%%%%%%%%%%%%%%%%%

%%% Include any author-defined commands here.
         
\newcommand{\BibTeX}{\rm B\kern-.05em{\sc i\kern-.025em b}\kern-.08em\TeX}

%%%%%%%%%%%%%%%%%%%%%%%%%%%%%%%%%%%%%%%%%%%%%%%%%%%%%%%%%%%%%%%%%%%%%%%%

%%% The following commands remove the headers in your paper. For final 
%%% papers, these will be inserted during the pagination process.

\pagestyle{fancy}
\fancyhead{}

%%% The next command prints the information defined in the preamble.

\maketitle

\vspace{-5pt}
\section{Introduction}
%background
Traffic congestion has emerged as a pressing issue in metropolises, leading to protracted travel and waiting time, heightened energy consumption, and diminished commuting comfort
\cite{levinson1998speed,tirachini2013estimation,economist2014cost,schrank20152015}.
Consequently, traffic signal control (TSC) has increasingly become a focal point of research, presenting an efficacious approach to alleviating such urban gridlock \cite{wei2019survey}.

Recently, the paradigm of TSC has predominantly shifted towards  deep reinforcement learning (RL) \cite{sutton2018reinforcement}. Such learning-based approaches \cite{li2016traffic,liang2018deep} can ``learn'' to give optimal actions directly based on the observation of intersections, which has proved its superiority over the conventional traffic-engineering-based methods such as SCATS and SCOOT \cite{roess2004traffic, hunt1982scoot,lowrie1990scats,gershenson2004self}, which are static models based on assumptions that could be unrealistic in front of the traffic dynamics.  
% traditional methods 需要假设，不是learningbase的
%Conventional traffic-engineering-based methods \cite{roess2004traffic, hunt1982scoot,lowrie1990scats,gershenson2004self} typically preconfigure control logic based on expert prior knowledge or select corresponding strategies based on observations of intersections. However, these methods are usually based on rigid assumptions such as uniform arrival rate that may not hold in real-life scenarios, thus lacking sufficient adaptability to cope with dynamic traffic situations.
% \hl{These methods tend to be shortsighted, struggling to account for the long-term impact on the system, and lacking the capacity for adaptive adjustment of coordinated strategies across multiple intersections.}
%Recently, the paradigm of TSC has predominantly shifted towards methods that utilize deep reinforcement learning (RL) %, with a specific emphasis on deep reinforcement learning
%\cite{sutton2018reinforcement}. These approaches \cite{li2016traffic,liang2018deep} learn to give optimal actions directly based on the observation of intersections, eliminating the need for unrealistic assumptions about control models. 
% Essentially, a TSC system based on RL first observes the traffic conditions, then generates and executes different actions (i.e., traffic control plans). Subsequently, it learns and adjusts strategies based on environmental feedback.
% RL-based solution: 分为两条技术路线，合作或多场景
Currently, there are two state-of-the-art solutions emerging in RL-based TSC: (1) cooperation among multi-agents and (2) learning via multi-scenarios.%, as concluded in Fig. \ref{fig:intro-two-sota}. 
%%\jiang{下面两个模块的顺序换成与title和方法里一致，先介绍weight再介绍co-train}

The first focuses on the cooperation of multiple intersections in one single \textit{scenario} (a simulation environment containing a set of intersections). With each intersection as an agent, multi-agent RL (MARL) \cite{mnih2016asynchronous,schulman2017proximal} have been developed. Most of the MARL-based TSC try to advocate cooperation by aggregating the information of the agents: they integrate the state of the target intersection with its neighboring intersections' states, either spatially \cite{van2016coordinated,wei2019colight} or spatiotemporally \cite{wang2020stmarl,wu2021dynstgat}, based on GNN \cite{wei2019colight} or GNN+LSTM/TCN to additionally capture long-range dependency \cite{wu2021dynstgat}, respectively. %, e.g., CoLight \cite{wei2019colight} used graph attention networks (GAT) \cite{velivckovic2017graph}. % with the objective of attaining coordinated control across multiple intersections.
Despite their potential, these approaches ignore that \textbf{how to coordinate with neighbors differs from scenarios}: training the model in only one particular scenario may lead to a local optimum due to overfitting, as discussed in \cite{jiang2023a}. 

%and do not possess an effective mechanism for extracting generalized dynamical information from diverse scenarios. 

The second type focuses on learning across multiple scenarios, such that the model could be more general for various regions or cities. 
To achieve this, different techniques such as meta RL \cite{zhang2020generalight,zang2020metalight}, attention mechanism \cite{oroojlooy2020attendlight} and standardization of intersections \cite{jiang2023a} have been proposed. For instance, MetaLight \cite{zang2020metalight} proposed a meta gradient learner through different datasets, and GESA \cite{jiang2023a} proposed a plug-and-play mapping module to enable multi-scenario co-training. These methods offer a potential solution to the overfitting problem mentioned before. 
However, they are all single-agent based, meaning one agent controls all the intersections, which may be an easy start for multi-scenario learning.
%从这里开始引出多场景协作的工作，然后陈述一下他的缺点：
MetaGAT \cite{lou2022meta} extended MetaLight to the multi-agent version by simply adding GAT in multi-scene training. 
However, these methods overlook that \textbf{how to utilize the unique knowledge of each scenario to facilitate the cooperation}: can we design an explicit mechanism for modeling experiential information within a single scenario?

\begin{figure}[t]
    \centering
    \includegraphics[width=0.95\columnwidth]{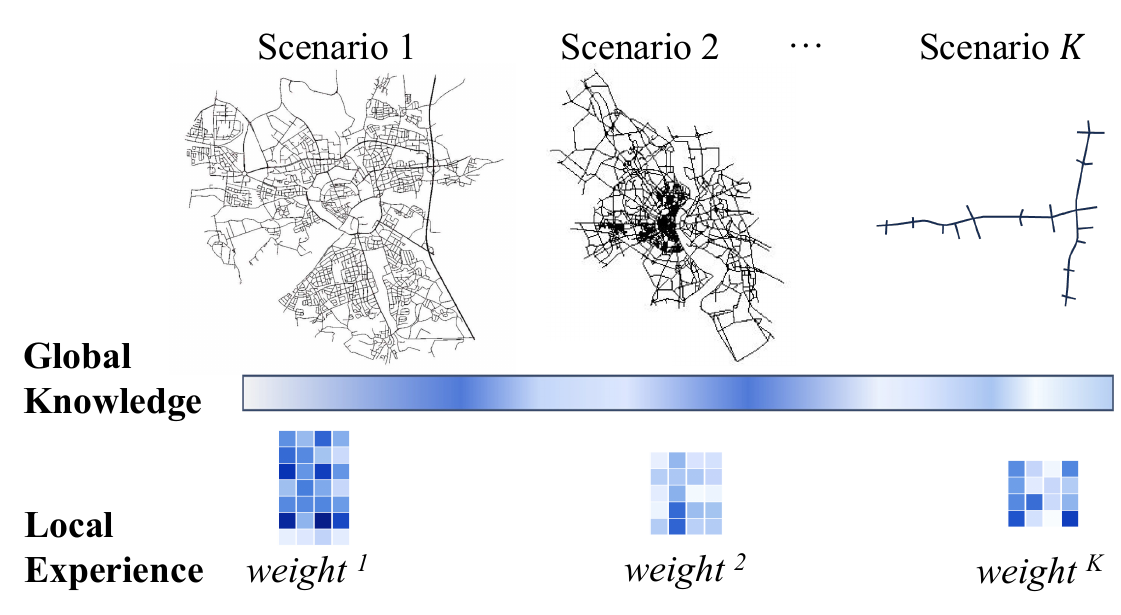}
    \caption{(1) Global knowledge from co-training, and (2) scenario-specific experiential knowledge formed as weights.}
    \label{fig:intro}
 % \vspace{-15pt}
\end{figure}
% \vspace{-10pt}

% \revise{To conclude, there is so far no work guiding how to coordinate multi-agent across multi-scenario. \hl{It is challenging since simply putting a multi-agent model in a multi-scenario learning setting could experience devastatingly unstable training, which has already taken shape in the single-agent version, i.e., MetaLight.} To encourage a long-term dynamic, introducing temporal modules such as LSTM is expected to aggravate the unstable learning, given the history is already unstable. To this end, this work is dedicated to solving the following research question: \textbf{how to coordinate multi-agent across multi-scenario stably and with the ability to reserve long-term experience}.}

%To address the drawbacks of these two types of models, 
To tackle these challenges, we propose \textbf{DuaLight} with two modules: \textbf{scenario-shared co-train} and \textbf{scenario-specific experiential weights}. %This innovative framework enables concurrent learning from both the universal dynamics information across multiple scenarios and the environmental experiential information within a single scenario. 
%We formulate the coordination parameters as the intersection-wise and feature-wise weights, guiding how to aggregate neighboring intersections' information and different observation features, respectively.
As shown in Fig \ref{fig:intro}: (1) \re{Some knowledge about the traffic underlining mechanisms is universal and commonly shared across various scenarios, for example, the flow is periodic, there are the potential morning/evening peak, and the merging and diversion of traffic flow along the network affect the traffic}. Co-Train module enables multi-scenario joint learning \re{of such global knowledge}. To encourage stability, only a \textbf{subset} of the model's parameters is trained concurrently, yet the essential coordination parameters are learned within each scenario. (2) \re{Some knowledge is scenario-specific: for example, each scenario has its unique distribution of the traffic flow, some tend to have more morning peaks and others more evening peaks, etc. This will affect different cooperation patterns from traffic lights.}  The experiential weight module defines the coordination parameters as the intersection-wise and feature-wise weights, guiding how to aggregate neighboring intersections' information and different observation features, respectively. The two weights are trained after observing a whole episode, capturing the long-term experiences. 

%This module can learn the environmental dynamic information about the neighbors and self-features within the scenario during training, which helps the model learn information that facilitates decision-making in a particular scenario. Specifically, this module can strengthen the experience information within a certain scenario by weighting the current neighbor observations and self-intersection observations.
%Additionally,  This module ensures the model's ability to extract the underlying dynamics information inherent across diverse scenarios.

The combination of these two modules encourages the model to learn and balance the shared dynamic information across multiple scenarios and the dynamic experiential information within a particular scenario, thus enabling the model to learn an effective representation to assist decision-making. This simple and effective dual design also supports our extension of using the neighbors \textbf{even from other scenarios}, which is a novel and promising discovery, as it can learn from similar intersections in other scenarios to further improve the ability of signal control.

%%\jiang{最后列一下贡献点} check
In summary, this paper has three main contributions:
\begin{itemize}
    \item To the best of our knowledge, we are the first that considers both scenario-common and scenario-specific information by co-train module and experiential weight module, respectively. This design also enables us to discover the potential of aggregating neighbors across scenarios. Overall, we coordinate multi-agents better across multi-scenario.
    \item Specifically, we design the scenario-specific experiential weights that encourage modeling the influence of neighbors and input features, adaptive to different scenarios. 
    \item We conduct experiments in both real-world and synthetic scenarios: DuaLight has the fastest and the most stable training and achieved the best results with 3-7\% improvements.
\end{itemize}
%First, the experiential weight module, which can learn the environmental dynamic information about neighbors and self-features within a particular scenario during training. Second, the Co-Train module, which enables the model to be trained across multiple scenarios, aims to learn the general dynamic information across different scenarios. Finally, we conduct experiments in both real-world and synthetic scenarios, where DuaLight achieved the best results or results close to the best across various metrics and scenarios.

\vspace{-10pt}

\section{Related Work}

\textbf{Learning to cooperate.}
In the realm of RL-based TSC, \cite{prashanth2010reinforcement,klingbeil2023centralized} directly train a centralized agent by using the observations of all intersections in a scenario as input to the model and providing a decision for each intersection. However, the complexity of these methods increases as the number of intersections increases, and it is hard to explore and optimize due to the curse of dimension in joint action space.
To ease this complexity, many MARL models take each intersection as an agent \cite{liu2023gplight}, 
% , 
with surrounding intersections considered for better decisions. 
% \cite{wiering2000multi} adds neighbors’ intersection states to the target intersection, while \cite{nishi2018traffic} adds neighbors’ hidden states. These methods treat the information from each neighbor equally. 
For example, CoLight \cite{wei2019colight} and MetaGAT~\cite{lou2022meta} employed GAT to assign varied weights to neighboring intersections. Yet, this approach primarily views neighboring intersections' information from a spatial standpoint in an instant short-sighted manner, without considering the influence of historical experiences on decision-making. To consider temporal information, STMARL \cite{wang2020stmarl} and DynSTGAT \cite{wu2021dynstgat} proposed to use LSTM or TCN to capture the historical state information, e.g., traffic flow, and employs GNN or GAT to extract the spatial dependencies. % among numerous relevant intersections.  applies TCN and GAT for handling historical spatial-temporal information while using GAT and LSTM for current spatial-temporal information management. 
% However, these methods only considered the temporal dependency of state, such as traffic flow; our experiential module instead captures the history of decision-making, i.e., the Markove decision process (MDP) dynamics via the gradients updated by Q-network. 
\revise{However, these methods only considered the temporal dependency of state (s1, s2, s3, ...), such as traffic flow; our experiential module instead allows for explicit capture of dynamic sequences (s1, a1, r1, s2, a2, r2, ...) through gradient propagation, offering a more comprehensive and nuanced understanding of scenario-specific decision dynamics.}
%\ziyue{our experiential info (MDP info) $\neq$ LSTM/TCN info (traffic flow info)}  
Moreover, these methods neglect that the impacts of their neighbors and input features are different across multiple scenarios. 
\revise{Our Experiential Weight uniquely adapts to varying scenario impacts through learnable weights. Unlike methods limited to single-scenario, DuaLight dynamically captures both unique and shared traffic dynamics across scenarios, enhancing adaptability and insight through backpropagation.}
% Our proposed Experiential Weight module instead is dedicated to learning for each scenario. 
%is adept at acquiring experiential information on \hl{self-features} \ziyue{all change to ``input feature''?} and the importance of neighbors within a certain scenario during the training process. This unique experiential knowledge in a certain scenario can integrate with the broad-ranging information gathered across various scenarios by the Co-Train module, thereby collaboratively bolstering the performance of the model.

\begin{figure}[t]
    \centering
    \includegraphics[width=0.99\columnwidth]{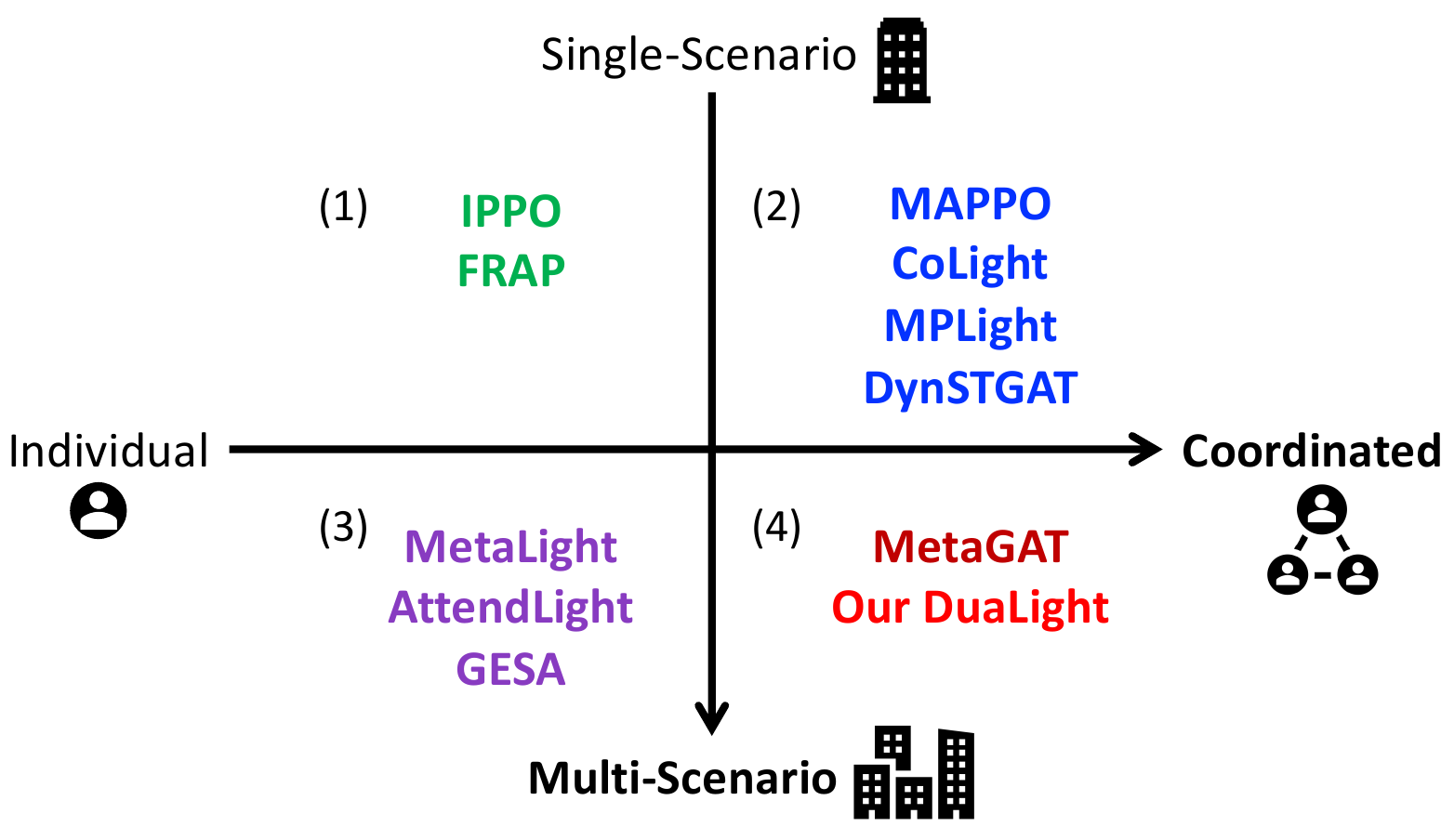}
    \caption{Two state-of-the-art methods: (1) whether it's multi-agent and (2) whether it's multi-scenario.}
    \label{fig:intro-two-sota}
% \vspace{-20pt}
\end{figure}
% \vspace{-10pt}

\textbf{Learning across multiple scenarios.}
% \ziyue{Reviewer 2: more detailed discussions on how it differs from or improves upon existing methods in terms of leveraging scenario-specific and scenario-shared knowledge. Maybe talk more about MetaLight's local function and meta-learner.}
Simultaneously, some methods examine training in multi-scenario for optimized performance. The single-agent version is mostly dominating. MetaLight \cite{zang2020metalight}, based on Meta RL \cite{finn2017model}, permitted joint scenario training and direct transfer to new scenarios. Similarly, GeneraLight \cite{zhang2020generalight} enhanced generalization by merging various traffic flows with generative adversarial networks and the MetaRL. AttendLight \cite{oroojlooy2020attendlight} introduced a training framework suitable for intersections of varying traffic flows and configurations by utilizing two attention models. The most recent GESA \cite{jiang2023a} presented a universal intersection normalization scheme and leveraged the A3C algorithm for joint training across multiple scenarios. However, a single agent controlling all is not optimal. Yet simply putting a multi-agent model in a multi-scenario learning setting could experience devastatingly unstable training, which has already taken shape in the single-agent MetaLight. MetaGAT {\cite{lou2022meta}}, as the most related work, tried to coordinate multi-agents across multi-scenario: It stabilized the training by separate task encoding (based on GAT+TCN) and controlling. %combined the GAT \cite{velivckovic2017graph} from CoLight into MetaLight framework. The similarity is that 
Similarly, we stabilize the learning by only training part of the parameters concurrently, and the scenario-specific experiential weights are learned separately. These weights not only explicitly preserve the long-term MDP experiences of each scenario, guiding a scenario-adaptive GAT and input-feature aggregation, but also enable a simple extension to use neighbors even across scenarios. %Although a direct comparison is not presented, we have compared the performance loss caused by not using the experience module in our ablation experiments.

%However, these methods jointly train all parameters within the model, which poses a challenge to preserving the intrinsic characteristics of a specific scenario. Additionally, these methods exhibit considerable fluctuations during the training process.
%% \jiang{1. Co-train并不是这里第一次提出的，可以下面先引一下GESA/AttendLight，后面再说一下GESA/AttendLight是所有参数都cotrain，很难保留场景自身的特性，这里我们提出了一种局部cotrain的module。2. 是Co-train module还是method,前后要统一。} check
%The Co-Train module that we propose is a simple yet effective method for targeted joint training of certain model parameters across multiple scenarios, thereby learning universal dynamic information from different scenarios.

%% Analogous to our Dual-experiential Weight Mechanism, several studies have leveraged self-attention mechanisms to enhance model performance. AttentionLight \cite{zhang2021attentionlight} ingests phase features as input, discerns the significance of each phase via the self-attention mechanism, and subsequently constructs new phase features grounded on each phase's relevance to inform its decisions. Nevertheless, these approaches solely consider features at the current time step, whereas our methodology emphasizes the longitudinal impact of the model's own features as well as the significance of its adjacent units.

\providecommand{\jiang}[1]{\textcolor{magenta}{#1}}

\section{Problem Definiton}
%% \jiang{有点口语化了（need to）} check
Before introducing the model, we shortly recap some key concepts integral to TSC. We recommend referring to \cite{jiang2023a} for more details.  %\ziyue{Why is a POMDP, not just an MDP?}.

\begin{figure}[t]
    \centering
    \includegraphics[width=0.99\columnwidth]{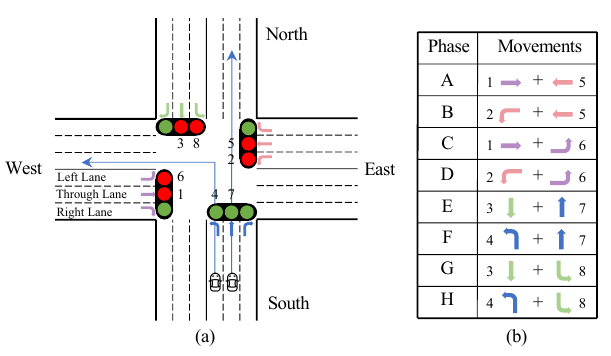}
    \caption{(a) A 4-arm intersection with eight traffic movements (1-8), presently controlled by phase F. (b) Eight phases (A-H), each having two non-conflicting traffic movements.}
    \label{fig:intersec-pahse}
\end{figure}

\begin{definition}[\textbf{Intersection}]
An \textbf{intersection} $I_i$ is where %a location in a traffic network that 
multiple roads connect and are controlled by a traffic light. A standard intersection, shown in Fig. \ref{fig:intersec-pahse}(a), consists of four \textbf{entrance arms} (N, S, E, W), each containing three possible entrance lanes: left-turn, through, and right-turn (also known as \textbf{traffic movements}). Each entrance arm has an exit arm as an outlet for vehicles. The majority of intersections are either with 3-arm or 4-arm structures.
\end{definition}

%\begin{definition}[\textbf{Entrance Arm}]
%Each entrance arm has a minimum of one entrance lane. As depicted in Figure \ref{fig:intersec-pahse}(a), a standard entrance arm consists of three entrance lanes, specifically right, straight through, and left.
%\end{definition}

%\begin{definition}[\textbf{Traffic Movement}]
%A traffic movement describes a vehicle's journey from an entrance lane, traversing the intersection, to an exit lane. A standard intersection %Figure\ref{fig:intersec-pahse}(a) 
%illustrates twelve movements, including left turns, right turns, and straight-through. Usually, turning-right movement is free from traffic control. Thus, in total, eight movements need TSC. %Practically, an entrance arm could accommodate multiple lanes dedicated to identical traffic movements.
%\end{definition}

\begin{definition}[\textbf{Phase}]
A traffic phase is a combination of traffic movements in which there is no conflict between them. Fig.\ref{fig:intersec-pahse}(b) depicts eight phases in a standard intersection. In the setting of RL in TSC, the action space $\mathcal{A}$ of the agent refers to \revise{select} the phase.
\end{definition}

%\begin{comment}
%% \jiang{这两个图还需要再优化一下（字体）} check
%% \jiang{加粗} check
%\end{comment}

We reconceptualize the TSC problem as a Partially Observable Markov Decision Process (POMDP) since each agent only observes part of the whole global state of the city.

\begin{problem}
%% \jiang{在这介绍一下MDP/POMDP， 下面几段介绍的有些繁琐了} check
\textbf{TSC as POMDP}: TSC is a sequential decision-making problem. Assuming there are $N$ intersections in a scenario, each intersection is controlled by an independent agent. The agent's goal is to learn a signal control policy to optimize travel time, which can be formulated as a POMDP
$\langle\mathcal{S}, \mathcal{O}, \mathcal{A}, \mathcal{P}, r, \gamma, \pi\rangle$. 
\end{problem}

\textbf{System state space $\mathcal{S}$} and \textbf{Partial observation space $\mathcal{O}$}: At the time $t$, each agent can observe a local observation $o_i^t \in \mathcal{O}$ from the global system state $s^t \in \mathcal{S}$, including the current phase and the number of stopped vehicles on the road. %% \jiang{确认一下这里面是不是只用了这两个state} check

\textbf{Action space $\mathcal{A}$}: $\mathcal{A}$ of each agent is to select one of the eight traffic phases (A-F) shown in Fig. \ref{fig:intersec-pahse}: Based on $o_i^t$, each agent selects an action $a_i^t \in  \mathcal{A}$ as the traffic signal control logic for the next time interval $\Delta t$. %% |jiang {使用图1及Definition 3.4进行一下解释。或者在Definition 3.4中说一下$\mathcal{A}$有哪些} check

\textbf{Transition probability $\mathcal{P}$}: A function for the system to enter the next state $s^{t+1}$, which is defined as $\mathcal{P}\left(s^{t+1} \mid s^t, a^t\right)$. This is an unknown function that encapsulates the dynamic information of traffic system operations.

\textbf{Reward function $r$}: After executing a phase, a reward can be obtained from the system based on the reward function. The immediate reward of agent $i$ at time $t$: $r_i^t = -w\sum_l q^t_{i,l}$, in which, $q^t_{i,l}$ represents the number of stopped vehicles on the approaching lane $l$ at time $t$, and $w$ represents the punishment coefficient, we
set it as 0.25. %\ziyue{$w=0.25$? since we already have $-$ in the reward} %% \jiang{确认一下这里面是不是只用了一个reward} check

%\textbf{Discount factor $\gamma$}: To ensure the long-term optimal traffic conditions, we set it as 0.95.

\textbf{Policy $\pi$}: a agent's controlling policy. At each time $t$, each agent follows policy $\pi(a^t \mid o^t)$ to make an action $a^t$ based on the current observation $o^t$, with the objective of minimizing all rewards $G_i^t = \sum_{t=\tau}^T \gamma^{t-\tau}r_i^t$, where $\gamma$ is the \textbf{discount factor} ($\gamma=0.95$).

\begin{figure*}[ht!]
  \centering
  \includegraphics[width=0.99\textwidth]{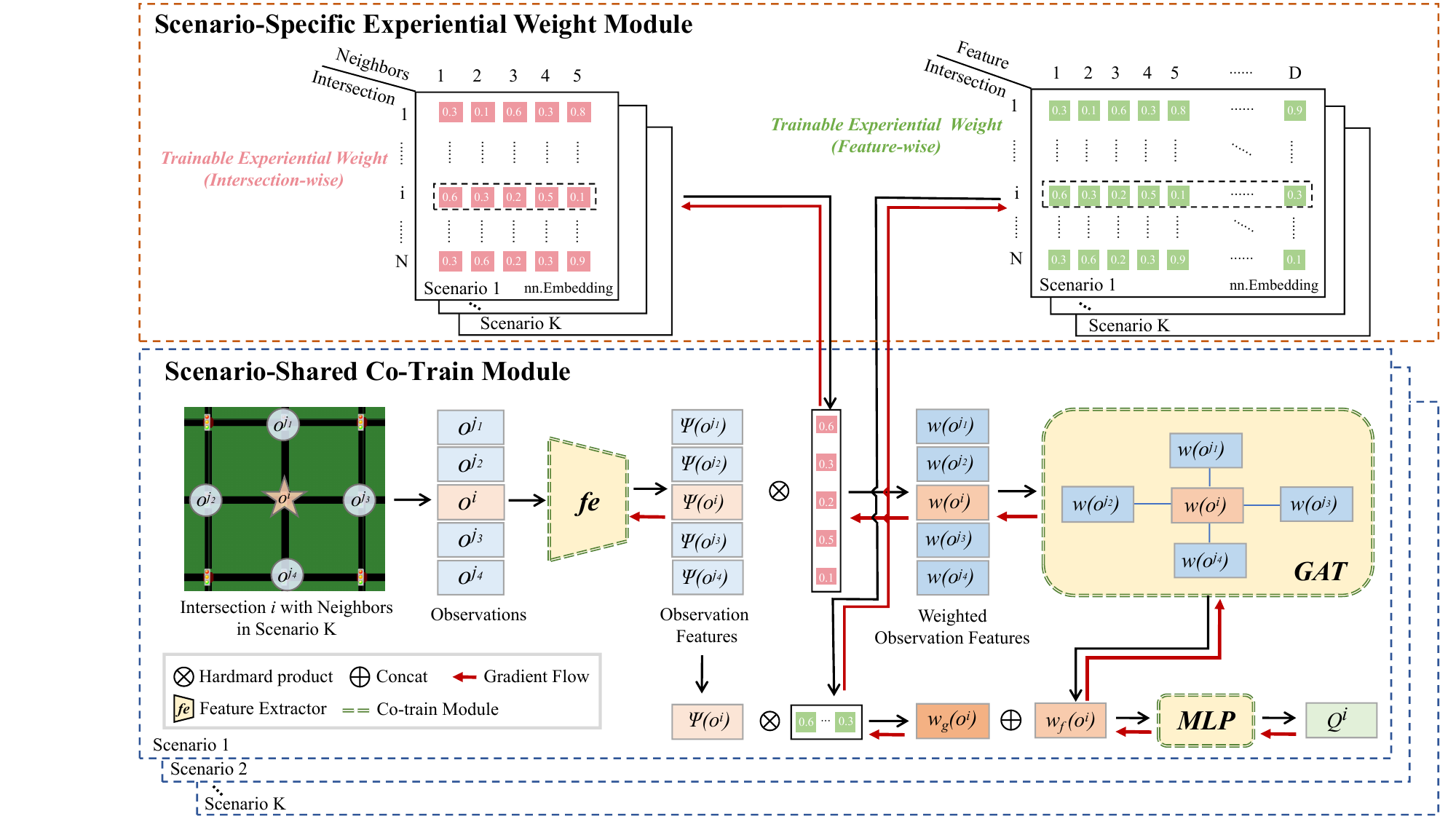}
  \caption{The illustration of our proposed DuaLight. The upper part of the figure represents the trainable experimental weights. The lower part gives the detailed interference process of all the modules.
  } %\ziyue{add ``experiential weight module'' and ``co-train module'' on the top of each block}}
  \label{fig:main_graph}
\vspace{-5pt}
\end{figure*}
\vspace{-5pt}

\section{Methodology}
\label{sec:method}
%% \jiang{看看能不能再整个Methodology中加入算法}
In the subsequent section, we delineate our proposed DuaLight, an end-to-end MARL architecture, as depicted in Fig.~\ref{fig:main_graph}. 
%% \jiang{分别介绍一下每个小段的内容，每个小段的动机} check
% Beginning with the feature extraction module, we proceed to the experiential weight module, emphasizing its information aggregation mechanism. We then describe the procedure for joint training across multiple scenarios, concluding with the prediction module and the training objective function. This breakdown aims to provide an overview of the operational dynamics of our architecture.
We will first introduce the feature extraction module, followed by the Scenario-Specific Experiential Weight Module and its coordination with GAT to aggregate observation information within neighbor intersections and self-features. Then, we will introduce the Scenario-Shared Co-train Module and the objective function for training.

\subsection{Feature Extractor}
In this stage, we first obtain feature representations from the simulator's raw observations of all lanes at the intersection, including the number of cars and the current stage. The multi-layer perceptron (MLP) is applied as the feature extractor $fe(\cdot)$ in the following.
\begin{equation}
    \psi(o^i) = fe(o^i)=\sigma(o^i \bm{W}_f+b_f),
\end{equation}
where $o^i \in \mathbb{R}^F$ is the raw observation of the intersection $i$ with the dimension of $F$, $\bm{W}_f \in \mathbb{R}^{F \times D}$, $b_f \in \mathbb{R}^D$ are the weight matrix and the bias of the MLP, and $\sigma(\cdot)$ denotes the $ReLU$ activation function. 

Thus, we obtain a $D$-dimensional representation as the base feature for each intersection. 
Next, we present the experiential weight to process these features further.

\subsection{Scenario-Specific Experiential Weight Module}

Merely projecting raw observations is often inadequate for traffic light control, as it requires a long-term and experiential understanding of both intersection-wise and self-feature-wise perceptions.
Specifically, the ability to perceive intersection-wise information is crucial for the decision-making process of an intersection, as it enables effective coordination between multiple intersections, leading to improved traffic flow and reduced congestion throughout the road network. Additionally, self-feature-wise perception is also essential for accurate decision-making that can alleviate congestion at the current intersection. Thus, we propose the experiential weight mechanism, which enables keeping the neighbor-intersection-wise and the self-feature-wise memory throughout the training process.

\subsubsection{\textbf{Two Trainable Experiential
Weights}}

Here, we define the trainable experiential
weight matrices as intersection-wise $\{Emb_{int}^k\}_{k=1}^{\revise{K}}$ 
% \ziyue{should also be 1 to $K$? look at fig-1, the pink matrices are also from senario-1 to scenario-K}
, and feature-wise $\{Emb_{fea}^k\}_{k=1}^K$, where for a given scenario $k$, $Emb_{int}^k \in \mathbb{R}^{N_k \times (1+N_{nei})}
$, $Emb_{fea}^k \in \mathbb{R}^{N_k \times D}$.
% \re{and $N = \max \{ N_k\}_{k=1}^K$}.
% Here, we define the trainable experiential
% weights (intersection-wise $Emb_{int}^k$ and feature-wise $Emb_{fea}^k$ in $k$-th scenario) as follows.
% \begin{equation}
%     \{Emb_{int}^k\}_{k=1}^N = nn.Embedding(N_k, 1+N_{nei}),
% \end{equation}
% \begin{equation}
%     \{Emb_{fea}^k\}_{k=1}^K =  nn.Embedding(N_k, D)
% \end{equation}
$K, N_k, N_{nei}$ denote the number of scenarios, the number of intersections in the scenario $k$, and the number of neighbors of an intersection, respectively. 
Moreover, to ensure precise attention to an agent's own features, it is crucial to assign a separate feature weight to each scenario $k$.
\revise{These two weight matrices are updated at each iteration during the training phase and are fixed during inference.}

In the implementation, the trainable experiential
weight matrix corresponds to a special MLP layer without bias and non-linear 
 activation function, implemented as a PyTorch module using $nn.Embedding$. 
 % \ziyue{what does it mean by ``without non-linear function''? }. 
During the training, these embeddings can be updated using the gradient $\bigtriangledown_{\phi }\mathcal{L}$ of optimization objective $\mathcal{L}$ (details in Eq.~(\ref{eq:nabla})) in an end-to-end manner, which enables these modules to contain intersection-wise and feature-wise representative historical information from the experiential replay buffer.
% , which contains representative historical messages.
%\ziyue{we could still add one equation about how the embedding is updated by the gradient}
\begin{equation}
    Emb_{\{int, fea\}}^k \leftarrow Emb_{\{int, fea\}}^k	- \alpha \nabla \mathcal{L}_{\phi}
\end{equation}

Next, we introduce the acquisition process of these weights, elaborated as follows.
% , illustrated as Figure~\ref{fig:acq_weight}.

% \begin{figure*}
%   \centering
%   \includegraphics[width=\textwidth]{figures/embedding.pdf}
%   \caption{The process of obtaining the dual-experiential weight.}
%   \label{fig:acq_weight}
% \end{figure*}

\paragraph{\textbf{1) The acquisition of intersection-wise experiential weight}}

To compute the intersection-wise experiential weight for each agent $i$, we begin by setting $N_{nei}=4$ \revise{and $\mathcal{N}_i = \{j_1, j_2, j_3, j_4\}$}. %and represent the IDs of its neighbors as $N_i = \{\mathcal{N}_1, \mathcal{N}_2, \mathcal{N}_3, \mathcal{N}_4\}$. 
Here, we assume that there are four neighbors that can be found through the nearest distance metric, and this assumption holds for the rest of the discussion. In Sec. {\ref{sec:cross-nei}}, we relax it by taking more neighbors, even from other scenarios.
To obtain the intersection-wise experiential weight, we use the operator $lookup(\mathbf{X}, i)$ to return the $i$-th row of $\mathbf{X}$:
\begin{equation}
weight_{int}^{k,i} = lookup(Emb_{int}^k, i) ,
    \label{eq:int}
\end{equation}
where $weight_{int}^{k,i} \in \mathbb{R}^{1+N_{nei}}$ and $Emb_{int}^k$ %is the trainable intersection-wise experiential weight for the scenario $k$, and it 
is trained in an end-to-end manner according to the RL target. Thus, as the training process progresses, the weight is endowed with high-level semantic information that represents the degree of attention between agents and their neighbors in the long run. This simple yet effective design (storable embedding + readout) allows a good extendability when even considering the neighbors from other scenarios (in Sec. {\ref{sec:cross-nei}}).

% We then define the lookup table operation as $lookup(Emb, ind)$, which indexes the elements with ID $ind$ from the table $Emb$. To obtain the intersection-wise experiential weight, we use this operator as follows:

% \begin{equation}
% weight_{int} = lookup(Emb_{int}, [i, \mathcal{N}_1, \mathcal{N}_2, \mathcal{N}_3, \mathcal{N}_4]) ,
% \end{equation}
% where $Emb_{int}$ denotes the trainable intersection-wise experiential weight, implemented as a PyTorch module using $nn.Embedding$, and it is trainable and updated in an end-to-end manner according to the reinforcement learning target.

\paragraph{\textbf{2) The acquisition of feature-wise experiential weight}}
\label{sec:fea_weight}
% 对于每一个场景来说，我们分别定义了该场景下的feature-wise experiential weight
For each scenario $k$ and each intersection $i$, we can obtain the feature-wise experiential weight as follows.
\begin{equation}
    weight_{fea}^{k,i} = lookup(Emb_{fea}^k, i)
        \label{eq:fea}
\end{equation}
where $weight_{fea}^{k,i} \in \mathbb{R}^{D}$. %and $Emb_{fea}^k$ denotes the trainable feature-wise experiential weight of the scenario $k$. After this operation, a weight vector with the same dimension of $D$ as the feature $\psi(o^i)$ is obtained.
Similarly, as the training process continues, the model learns the optimal weight for each intersection feature, allowing it to assign appropriate attention to each feature during decision-making. %By accurately estimating the current conditions, the learned weights enable the model to make more informed and precise decisions. 
The feature weight is critical to ensure the model adapting to varying scenarios and making reliable predictions, as each scenario may require a different emphasis on certain features.

\revise{
In summary, the model's two weights have distinct roles. The intersection-wise weight linearly transforms input features before the GAT layer, capturing complex feature relationships and improving the model's learning and generalization. The feature-wise weight ensures reliable predictions by adapting to different scenarios, each requiring emphasis on specific features.
}

Next, we will introduce how to integrate the experiential weights into the decision-making process.

% \ziyue{how frequently are these two weights updated? are they fixed in the inference? why do we need the two weights? and why they are better than other alternative solutions? e.g., what the intersection-wise weight essentially does is to linearly transform before the GAT attentions, i.e., an additional linear layer before GAT: why it is important and helpful?}

\subsubsection{\textbf{Scenario-Specific Knowledge Injection}}

% As shown in the lower part of Figure~\ref{fig:main_graph}, there involves multi-source message aggregation in the decision-making process.
As shown in the lower part of Fig.~\ref{fig:main_graph}, there involves scenario-specific knowledge injection in the decision-making process.

\paragraph{\textbf{1) The knowledge from the neighboring intersection}}

At the intersection-wise level, we take the experiential (global) and instant (local) impacts into consideration. The global experiential impact can be perceived through the intersection-wise weight, and the weighted observation feature of agent $i$ and its neighbors \re{$\mathcal{N}_i \in \mathbb{R}^{N_{nei}}$} can be represented as follows.
\begin{equation}
    w(o^i) = weight_{int}^i \otimes \psi(o^i)
\end{equation}
\begin{equation}
    \{w(o^j)\}_{j\in \mathcal{N}_i} = weight_{int}^j \otimes \psi(o^j), 
\end{equation}
where $\otimes$ denotes the Hardmard product.

Then we extract the local instant impact via GAT~\cite{velivckovic2017gat,wei2019colight} due to its powerful representation capacity.
Through the attention mechanism in GAT, the important coefficients $a^{ij}$ are computed:
\begin{equation} \small
    a^{ij} = \frac{{\exp (w(o^i)\bm{\widehat W}{{(w(o^j)\bm{\widehat W})}^{\mathrm{T}}}/\sqrt {{D'}} )}}{{\sum\limits_{\revise{l \in \mathcal{N}_i}} {\exp (w(o^i)\bm{\widehat W}{{(w(o^l)\bm{\widehat W})}^{\mathrm{T}}}/\sqrt {{D'}} )} }},\sum\limits_{j \in \mathcal{N}_i} {a^{ij}}  = 1,
    \label{eq:att}
\end{equation}
% \ziyue{$k$ should always index scenario.} 
where $l$ iterates among the set containing the agent $i$ and \revise{its neighbors $\mathcal{N}_i$}, $\bm{\widehat W} \in \mathbb{R}^{D \times D'}$ is a learnable weight matrix for the attention mechanism, $D'$ denotes the dimension of the latent vector.
% , and ${{\{ \mathcal{N}\} }^{ - i}}$ is the neighbors of the agent $i$. 
Moreover, multi-head attention (MHA) is used to stabilize the training process. 
%In essence, MHA allows the model to attend to different parts of the observation features using multiple attention mechanisms in parallel.
We apply the average pooling to the hidden vectors from each head and pass through a transformation to produce the final output.
Thus, the final latent feature $w_g(o^i)$ \re{aggregates $i$'s neighbors' information into $i$ by adopting GAT:}.
\begin{equation} 
w_g(o^i) = \sigma \left( {\frac{1}{M}\sum\limits_{m = 1}^M {\sum\limits_{j \in \mathcal{N}_i} {a_{m}^{ij}{\bm{W}^m}w(o^j)} } } \right),
\end{equation}
where $\sigma(\cdot)$ is the activation function, $M$ is the number of attention heads, $a_{m}^{ij}$ is the attention score of the $m$-th attention head in Eq. (\ref{eq:att}), $\bm{W}^m$ is the learnable matrix with respect to the head $m$.
% \ziyue{in Eq-6, the neighbors of }
% \ziyue{What is ${{\{ \mathcal{N}\} }^{ - i}}$. }

\paragraph{\textbf{2) The knowledge from the self input feature}}
At the feature-wise level, 
the feature $\phi(o^i)$ of the agent $i$ is weighted by the feature-wise weight from Sec.~\ref{sec:fea_weight}.2), calculated as follows.
\begin{equation}
    w_f(o^i) = weight_{fea}^{k,i} \otimes \phi(o^i)
\end{equation}
% \ziyue{what is $\otimes$?} 
At the end, the multi-source message is contacted and passed through an MLP layer to predict the final state-action value. 
\begin{equation}
\begin{aligned}
Q^i & = MLP(w_g(o^i) \oplus  w_f(o^i)) \\
& = (w_g(o^i) \oplus  w_f(o^i))\bm{W}^{o}+b^o
\end{aligned}
\label{eq:q^i}
\end{equation}
% \ziyue{what is $\oplus$?} 
where $p=8$ is the dimension of the action space, i.e., the number of phases. $\oplus$ denotes the concat operator, $\bm{W}^o \in \mathbb{R}^{2D \times p}$ and $b^o \in \mathbb{R}^p$ are the weight matrix and the bias vector of the output MLP layer.

\IncMargin{1em}
\begin{algorithm}[t] 
\footnotesize
\SetKwInOut{Ensure}{ensure}\SetKwInOut{Initialize}{initialize}
% \SetKwInOut{Output}{output}
	\Ensure{the co-training networks $C=\{fe, GAT, MLP\}$, and the experiential weights $Emb=\{Emb_{int}, Emb_{fea}\}$}
	\Initialize{$L$, $T$, $K$, $N_k$, $\mathcal{B}$; // The number of training episodes, the number of timesteps in an episode, the number of scenarios, the number of intersections in scenario $k$, and the experience replay buffer; } 
        \Initialize{the parameter $\theta$ for the co-training networks $C$, and $\phi$ for the experiential weights $Emb$; }
        % \Output{Optimized parameters $\theta_0$, $\theta_1$}
        % Initialize the network parameters of \textbf{fe, GAT and MLP} $\theta_0$  with random weight , the embedding parameters of \textbf{experiential weight} $\theta_1$ with weight 1\;
        % Initialize experience replay buffer $B$\;
        \For{episode $l = 1$ \KwTo $L$}{ 
        \For{timestep $t = 1$ \KwTo $T$}{ 
            \For{scenario $k = 1$ \KwTo $K$}{ 
                \For{intersection $i = 1$ \KwTo $N_k$}{ 
                    Observe $o^{k,i}_t$\;
                    Get the experiential weights by Eq.~(\ref{eq:int}), and Eq.~(\ref{eq:fea})\; 
                    Obtain action $a^{k,i}_t$ by $\arg\max Q^i$ of Eq.~(\ref{eq:q^i})\;
                    Receive the reward $r^{k,i}_t$ and the next observation $o^{k,i}_{t+1}$\;
                    Store the transition $(o^{k,i}_t, a^{k,i}_t, r^{k,i}_t, o^{k,i}_{t+1})$ in $\mathcal{B}$;\ 
                    % Sample a batch of samples from $B$\;
                }             
            Sample a minibatch $\mathcal{M}_1$ from $\mathcal{B}$\;
            Update the experiential weights $Emb$ using Eq.~(\ref{eq:theta_phi}a) with $\mathcal{M}_1$;\  
            } 
        }
        Sample a minibatch $\mathcal{M}_2$ from $\mathcal{B}$\;
        Update the co-train networks $C$ using Eq.~(\ref{eq:theta_phi}b) with $\mathcal{M}_2$;\
        }

 	 	  \caption{The Pseudo-code of DuaLight}
 	 	  \label{algo_disjdecomp1} 
 	 \end{algorithm}
 \DecMargin{1em}

%% \jiang{应该把training objective放到方法的最后} check
\subsection{Scenario-Shared Co-Train Module}
 %% \jiang{应该说与GESA一样，引入Co-train，不过这里我们不会让所有参数都进行Co-train} check

% \ziyue{We could write down a pseudo-algorithm for the training. By this algorithm, we could show that that what is updated thru $K$ scenario} 

%% \jiang{下两段来自于GESA，直接引用一下就行} check
To encourage the model to learn the common patterns that are independent of scenarios, we aim to co-train various scenarios together. %Some intersections is a 3-arm configuration with lanes accommodating identical traffic movements. 
The difficulties are: most RL-based TSC models assume homogeneity across intersections, i.e., equalities in observation space, action space, reward function, and policy $\pi$. However, the standard 4-arm intersection structure does not ubiquitously apply in the real world. There are 4-arm intersections with irregular angles and also 3-arm or 5-arm intersections. Different cities definitely have different intersections with different structures (with different numbers of entrance arms, different combinations of lanes). This observation necessitates the standardization of intersections, translating non-standard forms into a uniform 4-arm structure. %where each incoming lane accommodates a single predominant traffic movement.
To achieve this, we deploy a mapping approach in \cite{jiang2023a}: GESA uses the relative orientation of entrance arms (e.g., the relative angels) to decide the conflicting movements, instead of the absolute orientation (N, S, E, W). The ``missing'' entrance arm's corresponding state is masked, and its action is zero-padded. More detail is given in \cite{jiang2023a}.

%Non-4-arm intersections are transformed into a 4-arm structure through the introduction of virtual arms. 

%To enhance the training efficiency of RL algorithms and enable the learning of shared dynamical information across varying scenarios, we propose a Co-Train module that can be combined with DQN. 
Contrasting the approaches by GESA \cite{jiang2023a} and other MetaRL-based methods \cite{lou2022meta,yang2021meta,zang2020metalight}, where all parameters of the model undergo joint training across multiple scenarios, our Co-Train module only allows the parameters within the \textbf{fe, GAT}, and \textbf{MLP} in Fig. \ref{fig:main_graph} to participate in such joint training. Conversely, the Experiential Weight module employs data exclusively from a single scenario for their training. This mechanism enables the Experiential Weight module to concentrate more effectively on capturing information within a specific scenario, whereas the \textbf{fe, GAT}, and \textbf{MLP} modules focus on grasping the general information across various scenarios. % \revise{Besides, it also enhances the training efficiency and stability.}

%% \jiang{下面这一段是算法实现上的细节，感觉这里可以换一下，结合图2，说一下那些地方是co-train的就可以了，这些地方的参数是一样的。同是修改一下图2，Co-train的地方要明显一些，方便说明} check

%To accomplish this, it is necessary to standardize and align intersections across multiple scenarios, treating them as homogeneous agents with shared parameters. However, these agents will make distinct decisions based on the unique states they observe. 
During co-training, we use multi-processing, with each process using SUMO \cite{behrisch2011sumo} for interaction across different scenarios and subsequently aggregating all data from each process into a unified buffer. When we sample data from the buffer for model updating, data from different scenarios are sampled with equal weight. During the network update, the parameters of the three modules—\textbf{fe, GAT}, and \textbf{MLP}—are updated using data from all scenarios. %Conversely, the Experiential Weight module utilizes data from a single scenario for updating its corresponding parameters. This mechanism enables the Experiential Weight module to concentrate more effectively on capturing information within a specific scenario, while the \textbf{fe, GAT}, and \textbf{MLP} modules focus on grasping the general information across various scenarios.

Our results demonstrate that the incorporation of the Multi-Scenario Co-Train module expedites the convergence of the model. When paired with the Experiential Weight module, the model not only becomes more stable but also delivers improved performance.

\subsection{Training Objective}
\label{sec: q}
%In this section, we introduce how to optimize the overall algorithm.
We adopt the value-based reinforcement learning regime to define the loss. The parameter-sharing mechanism is applied across all the agents.
For scenario $k$, the objective is to find the optimal Q-function that maximizes the expected return.
\begin{equation}
    Q^k(s_t,a_t)=\mathbb{E}\left [ \sum_{t=0}^{\infty} \gamma r^k_{t} | s_t, a_t \right ],
\end{equation}
% \ziyue{what is ``$r_{t+i}$'', what is the $i$ here?} 
where $Q^k$ is the action-value function for the scenario $k$, $s_t$ and $a_t$ are the state and action at time step $t$, $r_t$ is the immediate reward received after taking action $a_t$, and $\gamma$ is the discounted factor.

At the time $t$, we can compute target Q value as below: 
\begin{equation}
\begin{matrix}
% Q^k(s_t, a_t) \leftarrow Q^k(s_t, a_t) + \alpha \cdot (Q^k_{target}(s_t, a_t)-Q^k(s_t, a_t)),
% \\
Q^k_{target}(s_t, a_t) = r_t^k + \gamma \max_{a'} Q^{k-}(s_{t+1}, a') ,
\end{matrix}
\label{eq:qvalue}
\end{equation}
where $Q^{k-}$ is the target network.% , $Q^k_{target}$ is the target Q value
This target network is a copy of the main network that is used to calculate the Q-values during training, but its parameters are not updated during the learning process. Instead, the target network's parameters are periodically updated with the parameters of the main network, which helps to stabilize the learning process and prevent oscillations or divergence.

Next, using Stochastic Gradient Descent (SGD) to approximate the gradient of Q-learning and compute the loss and its gradient, we can write down the following rules.
\begin{equation}
    \begin{matrix}
\mathcal{L} = \frac{1}{2} \| Q^k(s_t, a_t)-Q^k_{target}(s_t, a_t)\|^2,
 \\
\bigtriangledown_{\theta, \phi }\mathcal{L} = (Q^k(s_t, a_t)-Q^k_{target}(s_t, a_t))\bigtriangledown_{\theta, \phi }Q^k(s_t, a_t), 
\end{matrix}
\label{eq:nabla}
\end{equation}

% \begin{equation}
%     \begin{matrix}
% \theta \leftarrow \theta - \alpha \cdot \bigtriangledown_{\theta}\mathcal{L} , \
% \phi \leftarrow \phi - \alpha \cdot \bigtriangledown_{\phi }\mathcal{L} , \\
% \end{matrix}
% \label{eq:theta_phi}
% \end{equation}
Now we can update the parameters by 
\begin{equation}    
    % \begin{align}
        \theta \leftarrow \theta - \alpha \cdot \bigtriangledown_{\theta}\mathcal{L}, \ \ \
        \phi \leftarrow \phi - \alpha \cdot \bigtriangledown_{\phi }\mathcal{L} ,
    % \end{align}	
    \label{eq:theta_phi}
\end{equation}
% \end{subequations}
where $\alpha$ is the learning rate, $\theta$ denotes the parameter for the co-training networks $\{\textbf{fe},\textbf{GAT},\textbf{MLP}\}$, and $\phi$ denotes the experiential weights $Emb_{\{int, fea \}}$. The model is summarized in Algorithm \ref{algo_disjdecomp1}.

\begin{table*}[t]
\centering  
\resizebox{1\textwidth }{!}{%
\begin{tabular}{c|lllllll|lllllll}
\toprule
\multirow{2}{*}{\textbf{Methods}} 
& \multicolumn{7}{c|}{\textbf{Avg. Trip Time (seconds)}}        
& \multicolumn{7}{c}{\textbf{Avg. Delay Time (seconds)}}                                                       \\ \cline{2-15} 

& \multicolumn{1}{c}{\textit{Grid4$\times$4}} 
& \multicolumn{1}{c}{\textit{Grid5$\times$5}} 
& \multicolumn{1}{c}{\textit{Arterial4$\times$4}} 
& \multicolumn{1}{c}{\textit{Ingolstadt21}} 
& \multicolumn{1}{c}{\textit{Cologne8}} 
& \multicolumn{1}{c}{\textit{Fenglin}} 
& \multicolumn{1}{c|}{\textit{Nanshan}} 

& \multicolumn{1}{c}{\textit{Grid4$\times$4}} 
& \multicolumn{1}{c}{\textit{Grid5$\times$5}} 
& \multicolumn{1}{c}{\textit{Arterial4$\times$4}} 
& \multicolumn{1}{c}{\textit{Ingolstadt21}} 
& \multicolumn{1}{c}{\textit{Cologne8}} 
& \multicolumn{1}{c}{\textit{Fenglin}} 
& \multicolumn{1}{c}{\textit{Nanshan}} 
\\ \hline

FTC                               
& 206.68 \fontsize{6pt}{6pt}\selectfont{(0.54)}               
& 550.38 \fontsize{6pt}{6pt}\selectfont{(8.31)}               
& 828.38 \fontsize{6pt}{6pt}\selectfont{(8.17)}  
& 319.41 \fontsize{6pt}{6pt}\selectfont{(24.48)}      
& 124.4 \fontsize{6pt}{6pt}\selectfont{(1.99)}                  
& 344.76 \fontsize{6pt}{6pt}\selectfont{(6.84)}
& 729.02 \fontsize{6pt}{6pt}\selectfont{(37.03)}

& 94.64 \fontsize{6pt}{6pt}\selectfont{(0.43)}                
& 790.18 \fontsize{6pt}{6pt}\selectfont{(7.96)}               
& 1234.30 \fontsize{6pt}{6pt}\selectfont{(6.50)}     
& 183.70 \fontsize{6pt}{6pt}\selectfont{(26.21)}         
& 62.38 \fontsize{6pt}{6pt}\selectfont{(2.95)}    
& 283.13\fontsize{6pt}{6pt}\selectfont{( 12.78 )}
& 561.69 \fontsize{6pt}{6pt}\selectfont{(37.09)}
\\

MaxPressure                       
& 175.97 \fontsize{6pt}{6pt}\selectfont{(0.70)}               
& 274.15 \fontsize{6pt}{6pt}\selectfont{(15.23)}              
& 686.12 \fontsize{6pt}{6pt}\selectfont{(9.57)}   
& 375.25 \fontsize{6pt}{6pt}\selectfont{(2.40)}    
& 95.96 \fontsize{6pt}{6pt}\selectfont{(1.11)}             
& 316.01 \fontsize{6pt}{6pt}\selectfont{( 4.86 )}
& 720.89 \fontsize{6pt}{6pt}\selectfont{(29.94 )}

& 64.01 \fontsize{6pt}{6pt}\selectfont{(0.71)}                
& 240.00 \fontsize{6pt}{6pt}\selectfont{(18.43)}              
& 952.53 \fontsize{6pt}{6pt}\selectfont{(12.48)}   
& 275.36 \fontsize{6pt}{6pt}\selectfont{(14.38)}    
& 31.93 \fontsize{6pt}{6pt}\selectfont{(1.07)}                
& 372.08 \fontsize{6pt}{6pt}\selectfont{( 267.2 )}
& 553.94 \fontsize{6pt}{6pt}\selectfont{( 32.61 )}
\\ \hline

MPLight                           
& 179.51 \fontsize{6pt}{6pt}\selectfont{(0.95)}                 
& 261.76 \fontsize{6pt}{6pt}\selectfont{(6.60)}               
& 541.29 \fontsize{6pt}{6pt}\selectfont{(45.24)}  
& 319.28 \fontsize{6pt}{6pt}\selectfont{(10.48)}  
& 98.44 \fontsize{6pt}{6pt}\selectfont{(0.62)}               
& 329.81 \fontsize{6pt}{6pt}\selectfont{(4.19)}
& 668.81 \fontsize{6pt}{6pt}\selectfont{(7.92)}

& 67.52 \fontsize{6pt}{6pt}\selectfont{(0.97)}                
& \textbf{213.78 \fontsize{6pt}{6pt}\selectfont{(14.44)}}     
& 1083.18 \fontsize{6pt}{6pt}\selectfont{(63.38)}   
& 185.04 \fontsize{6pt}{6pt}\selectfont{(10.70)}  
& 34.38 \fontsize{6pt}{6pt}\selectfont{(0.63)}                
& 399.34 \fontsize{6pt}{6pt}\selectfont{(248.82)}
& 494.05 \fontsize{6pt}{6pt}\selectfont{(7.52)}
\\    
IPPO                              
& 167.62 \fontsize{6pt}{6pt}\selectfont{(2.42)}               
& 259.28 \fontsize{6pt}{6pt}\selectfont{(9.55)}               
& 431.31 \fontsize{6pt}{6pt}\selectfont{(28.55)}  
& 379.22 \fontsize{6pt}{6pt}\selectfont{(34.03)}  
& 90.87 \fontsize{6pt}{6pt}\selectfont{(0.40)}              
& 368.14 \fontsize{6pt}{6pt}\selectfont{(6.25)}
&  743.69  \fontsize{6pt}{6pt}\selectfont{(38.9)}            

& 56.38 \fontsize{6pt}{6pt}\selectfont{(1.46)}                
& 243.58 \fontsize{6pt}{6pt}\selectfont{(9.29)}               
& 914.58 \fontsize{6pt}{6pt}\selectfont{(36.90)}  
& 247.68 \fontsize{6pt}{6pt}\selectfont{(35.33)}   
& 26.82 \fontsize{6pt}{6pt}\selectfont{(0.43)}                
& 324.57 \fontsize{6pt}{6pt}\selectfont{(12.19)}
&  577.99 \fontsize{6pt}{6pt}\selectfont{(42.22)}
\\          

rMAPPO                            
& 164.96 \fontsize{6pt}{6pt}\selectfont{(1.87)}               
& 300.90 \fontsize{6pt}{6pt}\selectfont{(8.31)}               
& 565.67 \fontsize{6pt}{6pt}\selectfont{(44.8)}   
& 453.61 \fontsize{6pt}{6pt}\selectfont{(29.66)}   
& 97.68 \fontsize{6pt}{6pt}\selectfont{(2.03)}                
& 412.73 \fontsize{6pt}{6pt}\selectfont{(14.54)}
& 744.47 \fontsize{6pt}{6pt}\selectfont{(30.07)}            

& 53.65 \fontsize{6pt}{6pt}\selectfont{(1.00)}                
& 346.78 \fontsize{6pt}{6pt}\selectfont{(28.25)}              
& 1185.2 \fontsize{6pt}{6pt}\selectfont{(167.48)}  
& 372.2 \fontsize{6pt}{6pt}\selectfont{(39.85)}    
& 33.37 \fontsize{6pt}{6pt}\selectfont{(1.97)}                 
& 403.6 \fontsize{6pt}{6pt}\selectfont{(57.29)}
& 580.49 \fontsize{6pt}{6pt}\selectfont{(33.6)}
\\    

CoLight                           
& 163.52 \fontsize{6pt}{6pt}\selectfont{(0.00)}               
& {242.37 \fontsize{6pt}{6pt}\selectfont{(0.00)}}               
& 409.93 \fontsize{6pt}{6pt}\selectfont{(0.00)}  
& 337.46 \fontsize{6pt}{6pt}\selectfont{(0.00)}   
& \textbf{89.72 \fontsize{6pt}{6pt}\selectfont{(0.00)}}           
& 324.2 \fontsize{6pt}{6pt}\selectfont{(0.00  )}
& \textbf{608.01  \fontsize{6pt}{6pt}\selectfont{(0.00  )} }   

& 51.58 \fontsize{6pt}{6pt}\selectfont{(0.00)}                
& 248.32 \fontsize{6pt}{6pt}\selectfont{(0.00)}               
& 776.61 \fontsize{6pt}{6pt}\selectfont{(0.00)}    
& 226.06 \fontsize{6pt}{6pt}\selectfont{(0.00)}   
& {25.56 \fontsize{6pt}{6pt}\selectfont{(0.00)}}               
&  262.32 \fontsize{6pt}{6pt}\selectfont{( 0.00 )}
&  \textbf{428.95 \fontsize{6pt}{6pt}\selectfont{(0.00 )}}

\\ 

MetaLight                         
& 169.21 \fontsize{6pt}{6pt}\selectfont{(1.26)}               
& 265.51 \fontsize{6pt}{6pt}\selectfont{(10.53)}              
& 424.39 \fontsize{6pt}{6pt}\selectfont{(24.49)}  
& 349.89 \fontsize{6pt}{6pt}\selectfont{(2.65)}  
& 97.93 \fontsize{6pt}{6pt}\selectfont{(0.74)}              
& 316.57 \fontsize{6pt}{6pt}\selectfont{(4.29)}
& 653.23\fontsize{6pt}{6pt}\selectfont{(9.15)}                

& 57.56 \fontsize{6pt}{6pt}\selectfont{(0.76)}                
& 270.06 \fontsize{6pt}{6pt}\selectfont{(31.54)}             
& 873.28 \fontsize{6pt}{6pt}\selectfont{(39.01)}   
& 227.48 \fontsize{6pt}{6pt}\selectfont{(4.25)}    
& 29.01 \fontsize{6pt}{6pt}\selectfont{(0.69)}               
& 376.11  \fontsize{6pt}{6pt}\selectfont{( 244.85 )}
& 478.81 \fontsize{6pt}{6pt}\selectfont{(10.29)}
\\

MetaGAT                           
& 165.23 \fontsize{6pt}{6pt}\selectfont{(0.00)}   
& 266.60 \fontsize{6pt}{6pt}\selectfont{(0.00)}  
& \textbf{374.80 \fontsize{6pt}{6pt}\selectfont{(0.87)}}  
&  {290.73 \fontsize{6pt}{6pt} \selectfont{(0.45)}}   
& 90.74 \fontsize{6pt}{6pt}\selectfont{(0.00)}  
& \textbf{290.73 \fontsize{6pt}{6pt}\selectfont{( 0.45  )}}
& 676.42 \fontsize{6pt}{6pt}\selectfont{( 0.00 )}

& 53.20 \fontsize{6pt}{6pt}\selectfont{(0.00)}              
&  \underline{234.80 \fontsize{6pt}{6pt}\selectfont{(0.00)}}     
& {772.36 \fontsize{6pt}{6pt}\selectfont{(0.00)}}             
&  264.07 \fontsize{6pt}{6pt}\selectfont{(9.85)} 
&  26.85 \fontsize{6pt}{6pt}\selectfont{(0.00)}              
&  \textbf{176.86 \fontsize{6pt}{6pt}\selectfont{(2.37)}}
&   503.42 \fontsize{6pt}{6pt}\selectfont{(0.00)}                        
\\ \hline

DuaLight                            
& \textbf{{161.04 \fontsize{6pt}{6pt}\selectfont{(0.00)}}}   
& \textbf{{\textbf{221.83 \fontsize{6pt}{6pt}\selectfont{(0.00)}}}}       
& \underline{{{396.65\fontsize{6pt}{6pt}\selectfont{(0.00)}}}}           
& {\textbf{317.97 \fontsize{6pt}{6pt}\selectfont{(0.00)}}}    
& {{\underline{89.74 \fontsize{6pt}{6pt}\selectfont{(0.00)}}}}     
& \underline{313.22 \fontsize{6pt}{6pt}\selectfont{(4.88)}}
& \underline{609.89 \fontsize{6pt}{6pt}\selectfont{(0.00)}}

& \textbf{49.32 \fontsize{6pt}{6pt}\selectfont{(0.00)}}     
& {{237.71 \fontsize{6pt}{6pt}\selectfont{(0.00)}}}       
& {\textbf{756.99\fontsize{6pt}{6pt} \selectfont{(69.44)}}}   
& {\textbf{182.67 \fontsize{6pt}{6pt}\selectfont{(9.34)}}}   
& {\textbf{25.35 \fontsize{6pt}{6pt}\selectfont{(0.00)}}}     
&  \underline{260.87 \fontsize{6pt}{6pt}\selectfont{(0.00)}}
&  \underline{429.49 \fontsize{6pt}{6pt}\selectfont{(0.00 )}}

\\ \bottomrule
\end{tabular}%
}
\caption{
\revise{Performance of synthetic and real-world data, including the mean and standard deviation (in parentheses). Best results in boldface, and the second best results underlined.}
% Performance on synthetic and real-world data. \ziyue{what's the thing in ()}
}
\label{table: main table}
\vspace{-10pt}
\end{table*}

\section{Experiments}
In this section, we outline the configuration of our experiments, the dataset, the comparative methods, and design multi-dimensional experiments to verify the effectiveness of our proposed DuaLight.

% the evaluation metrics employed for the assessment of model performance. 
% Subsequently, we report the experimental findings derived from our proposed model in contrast with those from alternative algorithms. Additionally, we delve into an analysis of DuaLight by visualizing the acquired features through t-Distributed Stochastic Neighbor Embedding (t-SNE) and conducting ablation experiments to discern the respective roles of various modules.

\subsection{Experiment Settings}
\textbf{Environment Setting}: For performance evaluation, we adopt the  Simulation of Urban Mobility (SUMO)\footnote{https://www.eclipse.org/sumo/}, extensively acknowledged and embraced by both academia and industry, as our experimental simulation platform. Within this simulated framework, each individual simulation proceeds for a duration of 3600 seconds, with the model making its decisions at an interval of $\Delta t$ = 15 seconds. 

\textbf{Model Setting}: \revise{We provide the detailed hyper-parameter settings in Table~\ref{app:table_para} of Appendix~\ref{app:sec:hyper}.}
%Subsequent to a decision being finalized by the model, a 5-second yellow light phase precedes the transmission of this decision to the corresponding traffic signal for regulation.
\subsection{Datasets}
Our model is assessed using three synthetic datasets and four datasets derived from real-world scenarios, summarized in Table \ref{tab: data} of Appendix~\ref{app:datasets}. Synthetic Datasets include \textit{Grid $4 \times 4$} \cite{chen2020toward}, \textit{Avenue $4 \times 4$} \cite{ma2020feudal}, and \textit{Grid $5 \times 5$}. Real-world Datasets include \textit{Cologne8} \cite{varschen2006mikroskopische} and \textit{Ingolstadt21} \cite{lobo2020intas} from Germany, as well as \textit{Fenglin} and \textit{Nanshan} \cite{jiang2023a} from China. For more details, please refer to \cite{ault2021reinforcement,jiang2023a}.

\subsection{Compared Methods}

DuaLight is compared with two distinct categories of signal control models: the first category is traditional traffic-engineering-based models and the second is RL-based models.

\textbf{Traditional Methods:}
\begin{itemize}
\item \textbf{Fixed-timed Control (FTC)} \cite{roess2004traffic}: This method employs expert knowledge to manually assign fixed phase sequences and durations to each traffic signal.
\item \textbf{MaxPressure} \cite{varaiya2013max,kouvelas2014maximum}: The pressure at each intersection is estimated by gauging the number of vehicles and queue length. Subsequently, the algorithm invariably selects phases that maximize this pressure in a greedy manner.
\end{itemize}

\textbf{Reinforcement Learning-based Methods:}
\begin{itemize}
    \item \textbf{IPPO} \cite{ault2019learning,ault2021reinforcement}: In independent PPO agents, each traffic signal is modeled as an independent agent. They utilize the same network architecture, but their parameters are not shared.
    \item \textbf{MPLight} \cite{chen2020toward}: This algorithm is based on the phase competition FRAP framework \cite{zheng2019learning} and employs pressure as both state and reward for the DQN agents.
    \item \textbf{MetaLight} \cite{zang2020metalight}: It integrates the FRAP framework with Meta RL to facilitate swift adaptation to new scenarios and enhance overall performance. This algorithm bears similarities with our proposed Multi-Scenario Co-train module.
    \item \textbf{rMAPPO} \cite{schulman2017proximal,yu2022surprising}: It is a widely adopted MARL framework with an actor-critic architecture that leverages proximal policy optimization to boost the stability of training. In this instance, we employ a variant equipped with an RNN module to encode historical information.
    \item \textbf{CoLight} \cite{wei2019colight}: It utilizes a GAT to aggregate the state information of neighboring intersections.
    % \item \textbf{MetaGAT} \cite{lou2022meta}: It is also multi-agent multi-scenario method. Yet there is no code available. We implemented it as ablation of DuaLight without experiential weight module in Table \ref{table: Ablation Experiment}.
    \item \textbf{MetaGAT}~\cite{lou2022meta} utilizes GAT-based context to enhance collaborative interactions between intersections. It is also a multi-agent multi-scenario method.
\end{itemize}

%%\jiang{
%% 1. In line with 改成了 Consistent with
%% 2.  gauge the efficacy -> measure the efficacy of
%% 3. 不用介绍怎么实现的；可以换成介绍这三个指标
%% }

\subsection{Evaluation Metrics}
Consistent with \cite{ault2021reinforcement}, we utilize \textbf{Average Delay, Average Trip Time}, and \textbf{Average Waiting Time} as evaluation metrics to measure the efficacy of the various TSC models. Among them, Delay represents the delay caused by signalized intersections (stop or approach delay), Trip Time represents the total time for a vehicle to travel from its starting point to its destination, and Waiting Time represents the time spent by a vehicle waiting at the intersections.

% Within SUMO, we can access log files documenting vehicle trajectories to compute data pertinent to these aforementioned metrics. Specifically, the log files furnish details such as departure time, arrival time, and travel distance for each vehicle. Consequently, we can determine the Average Waiting Time, the Average Trip Time from commencement to completion, and the Average Delay engendered by congestion or other influencing factors as evaluation metrics for all vehicles engaged in a simulation.

%\vspace{-5pt}
\subsection{Main Results}
In this section, we introduce the results yielded by DuaLight and other methods based on the various evaluation metrics.

\begin{figure*}[ht!]
  \centering
  \includegraphics[width=0.99\textwidth]{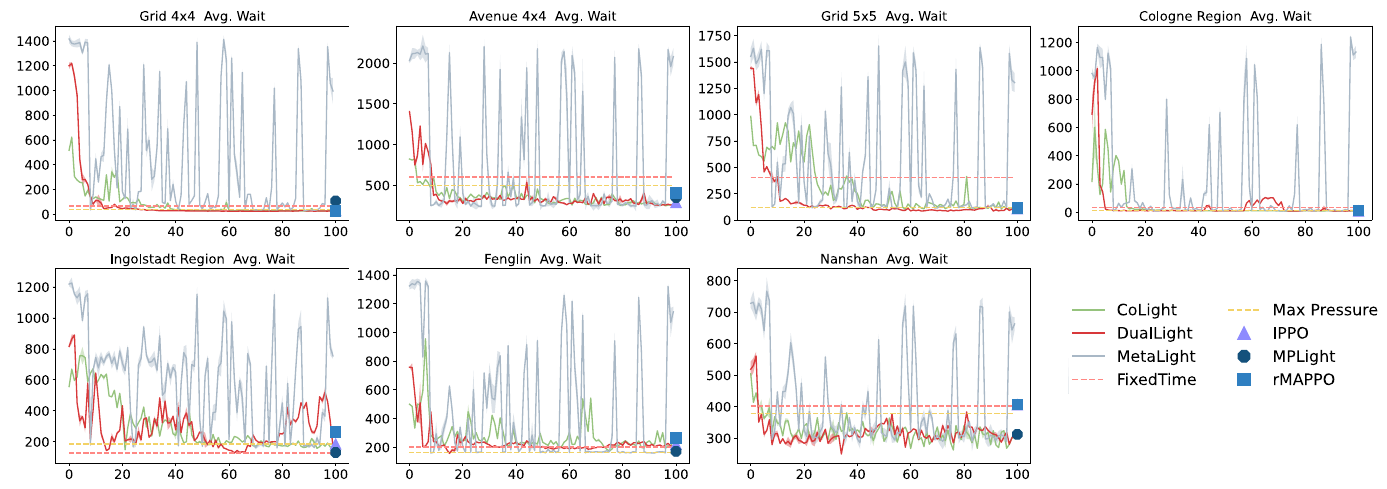}
  \caption{Learning curves of the Waiting Time metric over 5 random seeds}
  \label{fig:convergence}
\end{figure*}

\subsubsection{Overall Analysis}
Tables \ref{table: main table} %and \ref{tab: Improvement of DualLight} 
present the outcomes of our proposed DuaLight algorithm in comparison to other traditional control algorithms and RL-based algorithms. DuaLight exhibits optimal performance on the majority of indicators or holds competitive outcomes in relation to the best-performing algorithm.

% The results encapsulated in Table \ref{tab: Improvement of DualLight} exhibit a comparison between DualLight and the top-performing algorithm, excluding DualLight. These results emphasize the average improvement yielded by DualLight over other optimal results. 
We compare the performance of DuaLight with the best-performing algorithm. 
% Among all scenarios, DuaLight demonstrates the most substantial enhancement in the Average Waiting Time metric, with an average improvement of 4.52\%, 6.86\% on the synthetic dataset, and 2.76\% on the real-world scenario. 
Despite DuaLight achieving SOTA results on the Delay metric barring \textit{Grid $5 \times 5$} and \textit{Nanshan}, its average performance is not as exceptional due to its less impressive performance on \textit{Grid $5 \times 5$} compared to MPLight. Looking at the TripTime metric, \textit{Fenglin, Ingolstadt21}, and \textit{Grid $5 \times 5$} all achieved significant improvements, indicating that DuaLight can handle complex scenarios.
Full metrics evaluation is shown in Table~\ref{app:table: main table} of Appendix~\ref{app:main}.

\begin{table}[h]
\centering
\resizebox{0.99\columnwidth}{!}{
\begin{tabular}{llll}
\hline
Metric      & All  & Synthesized  & Real-world   \\ \hline
Avg. Delay      & -0.35\%          & -1.43\%          & 0.45\%       \\
Avg. Trip Time   & 2.03\%             & 4.41\%          & 0.25\%      \\
Avg. Wait      & 4.52\%                         & 6.86\%          & 2.76\%        \\ \hline
\end{tabular}
}
\caption{Improvement of DualLight across different scenarios}
\label{tab: Improvement of DualLight}
\end{table}

\subsubsection{Convergence Comparison}
Fig.~\ref{fig:convergence}~illustrates the learning curves of CoLight, MetaLight, and DuaLight with respect to the Average Waiting Time metric. The traditional control algorithms are represented by dashed lines, whereas other RL-based algorithms that require more training rounds to converge are marked with the best performance achieved. As evident from the graph, compared to MetaLight and CoLight, DuaLight's learning curve exhibits greater stability and has the best convergence speed in most scenarios. This indicates that the information from the experiential weight module and the shared information across different scenarios is fully utilized, enabling the model to achieve better performance and faster convergence speed: \revise{for example, in \textit{Grid $4 \times 4$}, DuaLight converges at step 10, whereas CoLight takes 30 steps; in \textit{Cologne8}, DuaLight converges at step 5, compared to CoLight's 20 steps.
Conversely, MetaLight has been consistently fluctuating.
% MetaLight instead has always been fluctuating.
} 
% \ziyue{Reviewer 3: The authors claim that the proposed DuaLight has the fastest training. However, there is no time complexity analysis, and no running time is reported, which makes the claims less convincing}

\subsection{Embedding Visualization of DuaLight}
\begin{figure}[t]
  \centering
  \includegraphics[width=\columnwidth]{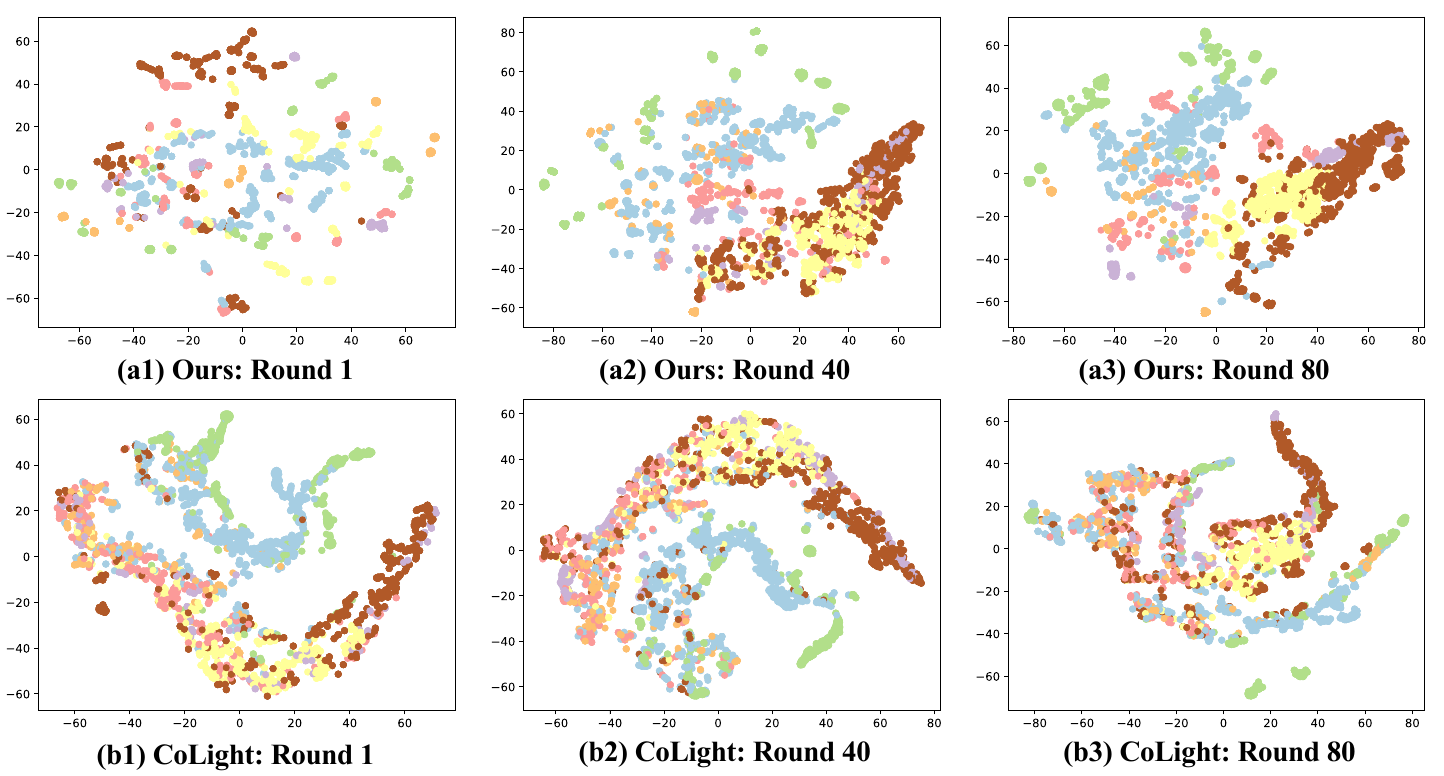}
  \caption{The t-SNE visualization of the hidden embeddings after message aggregation}
  \label{fig:TSNE}
\vspace{-10pt}
\end{figure}

To ascertain how our proposed scenario-specific experiential weight module is utilized, and to explore the information learned by it, we visualize the embeddings post message aggregation using t-SNE \cite{van2008visualizing} after different rounds. %More specifically, we utilize a trained model after a certain number of iterations for the simulation evaluation of all scenarios. 
We independently repeat the evaluation five times, and each time we select time steps 100-110 for visualization. For each agent, we extract its weighted embeddings before the MLP in Fig. \ref{fig:main_graph}, and visualize them via t-SNE, shown in Fig. \ref{fig:TSNE}.

A point in Fig. \ref{fig:TSNE} represents an agent, %after information weighting through the Intersection-wise and Feature-wise modules, visualized in a 2D plane.
and different colors represent agents from disparate scenarios.
% \ziyue{change the explanation according to the new Figure 5. also explain CoLight}
From Fig. \ref{fig:TSNE}(a1) to (a3), we observe that, after a certain number of training iterations, DuaLight's agent embeddings (weighted by experiential module) from the same scenario are moving closer to form one cluster. This suggests that our experiential weight module assists agents in capturing information within a certain scenario. While the embeddings weighted by GAT for CoLight, shown in Fig. \ref{fig:TSNE}(b1) to (b3), reveal that even after numerous training iterations, CoLight remains unable to distinguish differences between scenarios.

This can be attributed to our proposed learnable Intersection-wise and Feature-wise modules participating in each round of model updating. Through continuous iterations of model updating, these modules can aggregate historical and environmental experiential information and extract the dynamic characteristics of neighbors and self-feature information in the corresponding scenarios. 
More specifically, Intersection-wise can aid agents in comprehending the long-term impact of surrounding neighbors, while Feature-wise can assist agents in understanding the significance of different features. Conversely, CoLight predominantly focuses on local feature information, making it challenging to extract information about diverse scenarios.
\subsection{Ablation Studies}
To investigate the impact of each module on the overall performance of DuaLight, we conduct ablation experiments, including four settings: (1) without (w/o) the Co-Train module, (2) w/o the Experiential weight module, (3) w/o the Intersection-wise module only, and  (4) w/o the Feature-wise module only. 
%% \jiang{重要结果想展示的加粗}check

%\vspace{-6pt}
\begin{table}[ht]
    \centering
\resizebox{0.99\columnwidth}{!}{%
\begin{tabular}{cp{1.5cm}p{1.5cm}p{1.5cm}}
\hline
\diagbox{Model}{Metrics} & Delay & Trip Time & Wait \\
\hline
w/o  Co-Train & -4.85\%   & -1.55\%   & -4.58\%  \\ \hline
w/o Experiential weight &  -2.30\% &  -4.47\%  & -14.89\%  \\ \hline
w/o  Intersection-wise weight  & -5.53\%  & -1.61\%  & -6.71\%  \\ \hline
w/o  Feature-wise weight & -1.57\%   & -2.85\%   & -4.66\%  \\ \hline
\end{tabular}
}
\caption{The results of ablation experiments}  
    \label{table: Ablation Experiment}
\end{table}

\revise{
Table \ref{table: Ablation Experiment} presents the results. The last column ``Improve'' indicates the average percentage of performance improvement in the specific metric compared to the full DuaLight. Overall, the absence of any module will result in a decrease in model performance. 
We provide full metrics evaluation in Table~\ref{app:table: Ablation Experiment} of Appendix~\ref{app:abl}.
}

\section{Cross-scenario Neighbors}
\label{sec:cross-nei}
As mentioned before, our Co-Train + Experiential Weight design could easily be extended to even incorporate the ``neighbors'' from other scenarios. In this section, we provide some preliminary results and they are quite promising.% about injecting external knowledge about the cross-scenario neighbors.

\begin{figure}[t]
    \centering
    \includegraphics[width=\columnwidth]{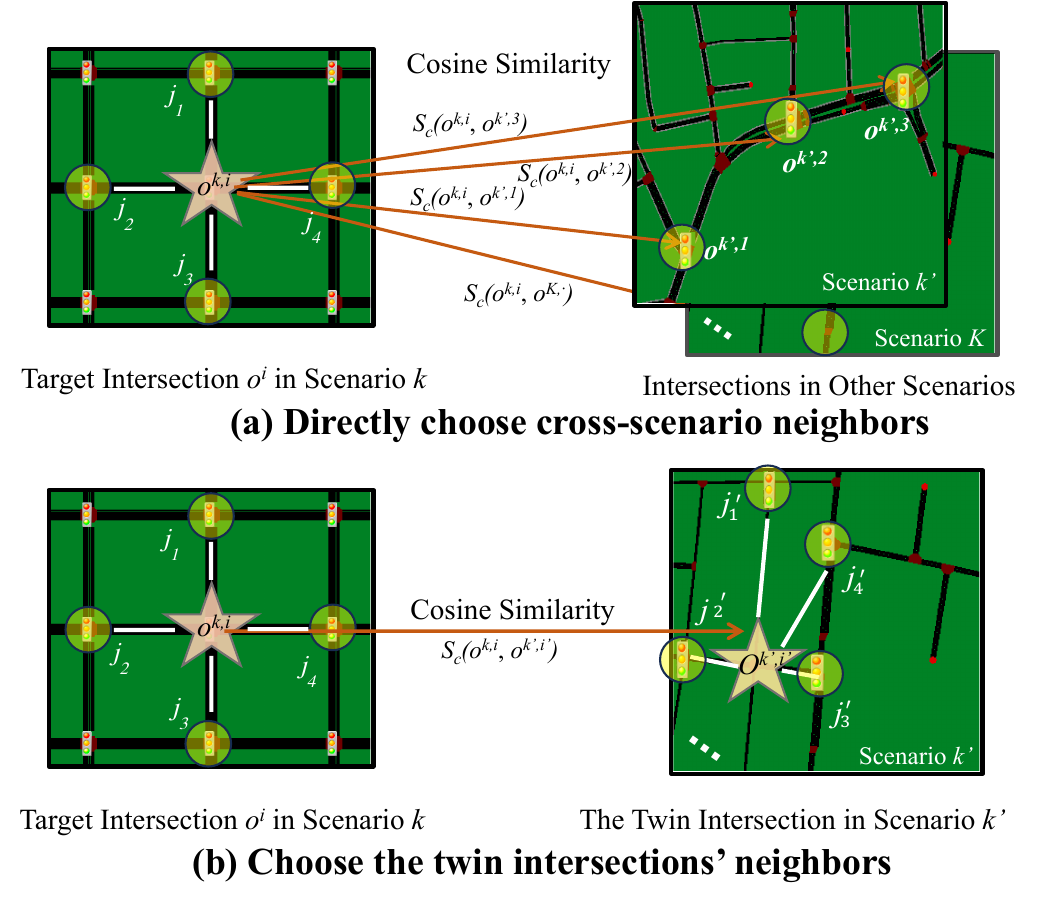}
    \caption{Two ways of picking cross-scenario neighbors}
    \label{fig:cos_sim}
\end{figure}
%\vspace{-5pt}

The illustration of how to select cross-scenario neighbors is shown in Fig.~\ref{fig:cos_sim}, elaborated as follows.
Given an observation of a target intersection $o^{k, i}$, we first compute the cosine similarity $S_c$ between \revise{the embedding of }$o^{k,i}$ and all the observations from other scenarios ${o^{k',\cdot}}$, where $k' \in \{1,...,K\} \backslash k$, as follows.
% \ziyue{$o^{k,i}$ is only the observation, how can we calculate their cos similarity? is it actually the embedding $\phi(o^{k,i})$}
\revise{
\begin{equation} \small
    S_c(\Psi(o^{k,i}), \Psi(o^{k',\cdot})) := \frac{\Psi(o^{k,i}) \cdot \Psi(o^{k',\cdot})}{ ||\Psi(o^{k,i})|| \ || \Psi(o^{k',\cdot})||}
\end{equation}
\vspace{2pt}
}

%\vspace{-6pt}
\begin{table}[t]
    \centering
\resizebox{0.99\columnwidth}{!}{%
\begin{tabular}{cp{1.5cm}p{1.5cm}p{1.5cm}}
\hline
\diagbox{Scenarios}{Metrics} & Delay & Trip Time & Wait \\
\hline
Grid $4 \times 4$ & \textbf{48.59 $\pm$ \footnotesize{0.0}} & \textbf{160.55 $\pm$ \footnotesize{0.0}} & \textbf{23.02 $\pm$ \footnotesize{0.0}} \\
\hline
Avenue $4 \times 4$ & \textbf{693.82 $\pm$ \footnotesize{0.0}} & 530.03 $\pm$ \footnotesize{7.6} & 372.06 $\pm$ \footnotesize{0.0} \\  
\hline
Grid $5 \times 5$ & \textbf{200.03 $\pm$ \footnotesize{0.0}} & \textbf{217.46 $\pm$ \footnotesize{0.0}} & \textbf{78.11 $\pm$ \footnotesize{0.0}} \\  
\hline
Cologne8 & 26.39  $\pm$ \footnotesize{0.0} & 90.71 $\pm$ \footnotesize{0.0} & \textbf{7.82 $\pm$ \footnotesize{0.0}} \\  
\hline
Ingolstadt21 & \textbf{168.1 $\pm$ \footnotesize{3.04}} & \textbf{282.1 $\pm$ \footnotesize{8.7}} & \textbf{95.87 $\pm$ \footnotesize{5.7}} \\  
\hline
Fenglin & \textbf{249.61 $\pm$ \footnotesize{0.0}} & 320.72 $\pm$ \footnotesize{0.0} & 171.02 $\pm$ \footnotesize{0.0} \\  
\hline
Nanshan & 530.23 $\pm$ \footnotesize{0.0} & 696.28 $\pm$ \footnotesize{0.0} & 379.14 $\pm$ \footnotesize{0.0} \\  
\hline
\end{tabular}
}
\caption{DuaLight++: with cross-scenario neighbors} 
\label{table:cross_scenarios}
\end{table}
\vspace{-15pt}

We design two ways of injecting cross-scenario neighbors. Way-(1) \textbf{the direct ones}: in Fig.~\ref{fig:cos_sim}(a), we select the messages of the top-$k$ (here $k=5$) correlated neighbors as the augmented external knowledge, or Way-(2) \textbf{the twin's}: in Fig.~\ref{fig:cos_sim}(b), we find the most similar neighbor (the twin) in scenario $k'$ and we use the twin and its four neighbors, together with the four neighbors from the same scenario 
(then in total $N_{nei} = 9$ 
% \ziyue{does $N_{nei}$ include itself or not?}
)
, we enhance the decision-making process.
Thus, benefiting from our framework design, we can directly get the weights from the intersection-wise weights with the least change: %and easily expand the message aggregation with cross-scenario neighbors. 
for Way-(1), we only need to re-train bigger intersection-wise weight matrice \revise{$weight_{int}^k \in \mathbb{R}^{N_k \times (1+N_{nei})}$ (here $1+N_{nei}=10$)}, thus capturing the weight of all the $N$ intersections in each scenario;
% \ziyue{is $N$ calculated from $\max$ or $\sum$?}
for Way-(2), we can directly use the current weights $weight_{int}^{k, i} \in \mathbb{R}^5$ and $weight_{int}^{k', i'} \in \mathbb{R}^5$ to concat as a 10-dimensional weight to embed the 9 neighbors to $o^{k,i}$.

As a preliminary study, we follow Way-(1) only, with the initial intersection-wise parameter set as 1. %but in reality, our intersection-wise module can be extended to multiple scenarios. 
Table~\ref{table:cross_scenarios}, compared with our DuaLight, demonstrates that incorporating the neighbors across scenarios based on similarity can help the model improve performance even further. %We believe that after using our intersection-wise weight, further performance is expected.

\section{Conclusion}
In this paper, we propose an effective RL approach and our method primarily consists of two integral components.
The first is the Experiential Weighted module, which supports the model in learning dynamic information about both the neighboring feature weights and self-feature weights within a specific scenario. When combined with the GAT Network, these two weights empower the model to focus simultaneously on real-time neighbor information and environmental information inherent in the scenario.
Secondly, we introduce the Co-train module, a component which is jointly trained with DQN across multiple scenarios. This facilitates the model's learning of shared and generic dynamic information across diverse scenarios.
%To assess the effectiveness of our proposed approach, we benchmark DuaLight against a variety of RL algorithms and traditional control methods, across multiple scenarios and metrics. 
Our results suggest that DuaLight delivers SOTA performance, or performs competitively against the existing SOTA results.
Embedding's visualization reveals that DuaLight is capable of learning superior feature representations, enabling better decisions. Moreover, we give the promising result of incorporating neighbors from other scenarios.

\revise{
\textbf{Limitations and Challenges}: The limitation is the requirement for retraining upon the addition of new scenarios. Future work will focus on developing a flexible weight learning mechanism to improve generalization to unseen scenarios, allowing for immediate adaptation without retraining. 
}

% \ziyue{Reviewer 2: the article could provide more information on potential challenges or limitations faced during the implementation of the DuaLight}

%\clearpage
\bibliographystyle{unsrt}
\bibliography{reference}

\clearpage
% \appendix
\onecolumn
\renewcommand*\appendixpagename{\centering Appendices}
\begin{appendices}
\setcounter{figure}{0}
\setcounter{section}{0}
\renewcommand{\thefigure}{A\arabic{figure}}
\setcounter{table}{0}
\renewcommand{\thetable}{A\arabic{table}}

\setcounter{secnumdepth}{2}

\section{Detailed data statistics of datasets}

\label{app:datasets}

Table~\ref{tab: data}~presents data statistics for different datasets. These datasets include traffic intersections from various countries and types. For simpler scenarios like Grid 4 $\times$ 4 and Grid 5 $\times$ 5, they show relatively simple structures. In contrast, complex scenarios such as Ingolstadt21 and Nanshan exhibit a larger number of intersections with more intricate structures. The intersections in the datasets are categorized as 2-arm, 3-arm, and 4-arm, indicating the number of exits in each intersection. These statistics provide a foundation for the subsequent sections, which involve model evaluation and result analysis.

\begin{table}[ht]
\centering
\resizebox{0.7\columnwidth}{!}{%
\begin{tabular}{lllllll}
\hline
Dataset      & Country & Type & Total Int. & 2-arm & 3-arm & 4-arm \\ \hline
Grid $4 \times 4$   & virtual & region & 16                         & 0          & 0          & 16         \\
Avenue $4 \times 4$   & virtual & region & 16                         & 0          & 0          & 16         \\
Grid $5 \times 5$     & virtual & region & 25                         & 0          & 0          & 25         \\
Cologne8     & Germany & region & 8                          & 1          & 3          & 4          \\
Ingolstadt21 & Germany & region & 21                         & 0          & 17         & 4          \\
Fenglin      & China & corridor & 7                          & 0          & 2          & 5          \\
Nanshan      & China & region & 29                         & 1          & 6          & 22         \\ \hline
\end{tabular}
}
\caption{Data statistics of datasets}
\label{tab:Data statistics of datasets}
\label{tab: data}
\end{table}

\section{Detailed hyperparameter settings }
\label{app:sec:hyper}
In this section, we provide a comprehensive outline of the hyperparameter configurations used in our experiments, facilitating result replication and understanding of our model's performance. 

\begin{table}[ht!]
    \centering
    \begin{tabular}{cc}
           \toprule
           Description &  Value \\
           \midrule
           optimizer       &    Adam   \\
        learning rate $\alpha$   &     $0.001$  \\
        $\epsilon$       &    $0.8$  \\
        $\epsilon_{min}$       &    $0.1$  \\
        $\epsilon_{decay}$       &    $0.999$  \\
        discounted factor $\gamma$   &    $0.95$  \\
        seed             &   $[0,10)$  \\
        total epochs $L$   &     $200$  \\
        total time steps in one episode $T$   &     $240$  \\
        batch size     &     $20$  \\
        eval interval  &     $1$  \\
        eval episodes  &     $100$   \\
        number of scenarios $K$          &   $7$  \\
        number of neighbors $N_{nei}$          &   $4$  \\
        dimension of raw observations $F$          &   $16$  \\
        dimension of features $D$          &   $16$  \\
        number of attention heads $m$          &   $1$  \\
        hidden size of GAT $D'$          &   $32$  \\
        hidden size of MLP layer $p$          &   $32$  \\
           \bottomrule
    \end{tabular}
    \caption{\revise{The hyper-parameters of Dualight. }}
    \label{app:table_para}
\end{table}

% \ziyue{Reviewer 3: The details of experimental settings are missing. It is not clear the number of layers, dimensions, etc.} 

\section{Additional Main Results}
\label{app:main}
In this section, we present the results of the complete metric evaluation of the main experiments.

% We compare the performance of DuaLight with the best-performing algorithm other than DuaLight. 
Among all scenarios, DuaLight demonstrates the most substantial enhancement in the Average Waiting Time metric, with an average improvement of 4.52\%, 6.86\% on the synthetic dataset, and 2.76\% on the real-world scenario. 
% Despite DuaLight achieving SOTA results on the Delay metric barring \textit{Grid $5 \times 5$} and \textit{Nanshan}, its average performance is not as exceptional due to its less impressive performance on \textit{Grid $5 \times 5$} compared to MPLight. Looking at the TripTime metric, \textit{Fenglin, Ingolstadt21}, and \textit{Grid $5 \times 5$} all achieved significant improvements, indicating that DuaLight can handle complex scenarios.

\begin{table*}[ht!]
    \centering
   \footnotesize 
    \resizebox{\textwidth}{!}{%
\begin{tabular}{l@{\hspace{0.7em}}l@{\hspace{0.7em}}l@{\hspace{0.7em}}l@{\hspace{0.7em}}l@{\hspace{0.7em}}l@{\hspace{0.7em}}l@{\hspace{0.7em}}l@{\hspace{0.7em}}l}
\hline
Model & Metric & Grid $4 \times 4$ & Avenue $4 \times 4$ & Grid $5 \times 5$ & Cologne8 & Ingolstadt21 & Fenglin & Nanshan \\ \hline
\multirow{3}{*}{FTC} & Delay & 94.64 $\pm$ 0.43 & 1234.3 $\pm$ 6.5 & 790.18 $\pm$ 7.96 & 62.38 $\pm$ 2.95 & 183.70 $\pm$ 26.21 & 283.13 $\pm$ 12.78 & 561.69 $\pm$ 37.09 \\
 & Trip Time & 206.68 $\pm$ 0.54 & 828.38 $\pm$ 8.17 & 550.38 $\pm$ 8.31 & 124.4 $\pm$ 1.99 & 319.41 $\pm$ 24.48 & 344.76 $\pm$ 6.84 & 729.02 $\pm$ 37.03 \\
 & Wait & 66.12 $\pm$ 0.32 & 599.1 $\pm$ 6.4 & 408.25 $\pm$ 7.09 & 35.3 $\pm$ 1.05 & 127.31 $\pm$ 20.06 & 200.02 $\pm$ 6.38 & 388.77 $\pm$ 26.33 \\ \hline
 
\multirow{3}{*}{MaxPressure} & Delay & 64.01 $\pm$ 0.71 & 952.53 $\pm$ 12.48 & 240.0 $\pm$ 18.43 & 31.93 $\pm$ 1.07 & 275.36 $\pm$ 14.38 & 372.08 $\pm$ 267.72 & 553.94 $\pm$ 32.61 \\
 & Trip Time & 175.97 $\pm$ 0.7 & 686.12 $\pm$ 9.57 & 274.15 $\pm$ 15.23 & 95.96 $\pm$ 1.11 & 375.25 $\pm$ 2.4 & 316.01 $\pm$ 4.86 & 720.89 $\pm$ 29.94 \\
 & Wait & 37.78 $\pm$ 0.65 & 495.52 $\pm$ 10.52 & 116.85 $\pm$ 12.5 & 11.19 $\pm$ 0.44 & 184.97 $\pm$ 3.56 & 161.78 $\pm$ 4.43 & 379.23 $\pm$ 23.55\\ \hline
 
\multirow{3}{*}{IPPO} & Delay & 56.38 $\pm$ 1.46  & 914.58 $\pm$ 36.9 & 243.58 $\pm$ 9.29 & 26.82 $\pm$ 0.43 & 247.68 $\pm$ 35.33 & 324.57 $\pm$ 12.19 & 577.99 $\pm$ 42.22 \\
 & Trip Time & 167.62 $\pm$ 2.42 & 431.31 $\pm$ 28.55 & 259.28 $\pm$ 9.55 & 90.87 $\pm$ 0.4 & 379.22 $\pm$ 34.03 & 368.14 $\pm$ 6.25 &  743.69 $\pm$ 38.9 \\
 & Wait & 29.59 $\pm$ 1.15 & 289.89 $\pm$ 25.53 & 108.61 $\pm$ 8.75 & 8.47 $\pm$ 0.2 & 185.88 $\pm$ 32.55 & 222.24 $\pm$ 8.85 & 405.84 $\pm$ 26.73 \\ \hline
 
 \multirow{3}{*}{MPLight} & Delay & 67.52 $\pm$ 0.97  & 1083.18 $\pm$ 63.38 & \textbf{213.78$\pm$ 14.44}  &34.38$\pm$0.63  & 185.04 $\pm$ 10.75 &399.34$\pm$248.82  &494.05 $\pm$ 7.52  \\
 & Trip Time &179.51$\pm$0.95  & 541.29 $\pm$ 45.24 & 261.76$\pm$ 6.60 &98.44$\pm$0.62  & 319.28 $\pm$ 10.48 &329.81$\pm$4.19  &668.81 $\pm$ 7.92   \\
 & Wait &41.14$\pm$0.79  & 349.69 $\pm$ 43.48 & 106.82$\pm$ 4.26 &12.06$\pm$0.32  & 130.57 $\pm$ 9.31 &171.24$\pm$4.76  &311.93 $\pm$ 6.93 \\ \hline
 
 \multirow{3}{*}{MetaLight} & Delay & 57.56 $\pm$ 0.76  & 873.28 $\pm$ 39.01 & 270.06$\pm$ 31.54  &29.01$\pm$0.69  & 227.48$\pm$ 4.25 &376.11$\pm$244.85  &478.81$\pm$10.29  \\
 & Trip Time &169.21$\pm$1.26  & 424.39 $\pm$ 24.49 & 265.51$\pm$ 10.53 &97.93$\pm$0.74  & 349.89 $\pm$ 2.65 & 316.57$\pm$4.29  &653.23 $\pm$ 9.15  \\
 & Wait &31.81$\pm$0.34  & \textbf{241.88 $\pm$ 11.29} & 114.49$\pm$ 2.59 &8.48$\pm$0.24  & 160.05 $\pm$ 3.09 & 160.38$\pm$3.09  &285.25 $\pm$ 18.59  \\ \hline
 
\multirow{3}{*}{rMAPPO} & Delay & 53.65 $\pm$ 1.0  & 1185.2 $\pm$ 167.48 & 346.78 $\pm$ 28.25 & 33.37 $\pm$ 1.97 & 372.2 $\pm$ 39.85 & 403.6 $\pm$ 57.29 & 580.49 $\pm$ 33.6 \\
 & Trip Time & 164.96 $\pm$ 1.87 & 565.67 $\pm$ 44.8 & 300.9 $\pm$ 8.31 &  97.68 $\pm$ 2.03 & 453.61 $\pm$ 29.66 & 412.73 $\pm$ 14.54 & 744.47 $\pm$ 30.07 \\
 & Wait & 27.07 $\pm$ 0.54 & 407.25 $\pm$ 41.15 & 153.38 $\pm$ 7.44 & 14.01 $\pm$ 1.12 & 264.41 $\pm$ 28.36 & 265.79 $\pm$ 14.82 & 407.31 $\pm$ 32.02 \\ \hline
 
\multirow{3}{*}{CoLight} & Delay & 51.58 $\pm$ 0.00 & 776.61 $\pm$ 0.00 & 248.32 $\pm$ 0.00 & 25.56 $\pm$ 0.00 & 226.06 $\pm$ 0.00 & 262.32 $\pm$ 0.00 & \textbf{428.95 $\pm$ 0.00} \\
 & Trip Time & 163.52 $\pm$ 0.00 & 409.93 $\pm$ 0.00 & 242.37 $\pm$ 0.00 & \textbf{89.72 $\pm$ 0.00} & 337.46 $\pm$ 0.00 & 324.2 $\pm$ 0.00 & \textbf{608.01 $\pm$ 0.00} \\
 & Wait & 25.56 $\pm$ 0.00 & 247.88 $\pm$ 0.00 & 100.58 $\pm$ 0.00 & 8.18 $\pm$ 0.00 & 147.91 $\pm$ 0.00 & 173.18 $\pm$ 0.00 & 262.45 $\pm$ 0.00 \\ \hline

\multirow{3}{*}{{MetaGAT}} & Delay & 53.20   $\pm$ 0.00 & {772.36   $\pm$ 0.00} & 234.80   $\pm$ 0.00 & 26.85   $\pm$ 0.00 &  264.07 $\pm$  9.85  &  \textbf{176.86  $\pm$  2.37} &  503.42   $\pm$ 0.00 \\
 & Trip Time & 165.23   $\pm$ 0.00 & \textbf{374.80   $\pm$ 0.00} & 266.60   $\pm$ 0.00 & 90.74   $\pm$ 0.00 & 
 319.44  $\pm$ 2.75   & \textbf{290.73  $\pm$  0.45} & 676.42   $\pm$ 0.00 \\
 & Wait & 27.25   $\pm$ 0.00 & 219.36   $\pm$ 0.27 & 123.78   $\pm$ 0.00 & 8.18   $\pm$ 0.00 & 164.12 $\pm$ 4.2  & \textbf{103.35   $\pm$ 0.62} & 325.17   $\pm$ 0.00 \\ 
 \hline

\multirow{3}{*}{\begin{tabular}[c]{@{}l@{}}DuaLight\end{tabular}} & Delay & \textbf{49.32 $\pm$ 0.00} & \textbf{756.99 $\pm$ 69.44} & 237.71 $\pm$ 0.00 & \textbf{25.35 $\pm$ 0.00} & \textbf{182.67 $\pm$ 9.34} & {260.87 $\pm$ 0.00} & 429.49 $\pm$ 0.00 \\
 & Trip Time & \textbf{161.04 $\pm$ 0.00} & {396.65 $\pm$ 0.00} & \textbf{221.83 $\pm$ 0.00} & 89.74 $\pm$ 0.00 & \textbf{317.97 $\pm$ 0.00} & {313.22 $\pm$ 4.88} & 609.89 $\pm$ 0.00 \\
 & Wait & \textbf{23.49 $\pm$ 0.00} & 252.43 $\pm$ 3.61 & \textbf{83.64 $\pm$ 0.00} & \textbf{7.85 $\pm$ 0.17} & \textbf{126.91 $\pm$ 0.00} & {157.53 $\pm$ 4.15} & \textbf{249.56 $\pm$ 0.00} \\ \hline
\end{tabular}
}
    \caption{Performance on synthetic data and real-world data}
    \label{app:table: main table}
\end{table*}

\section{Additional Ablation Results}

\label{app:abl}

In this section, we present the results of the complete metric evaluation of the ablation study.
In particular, the lack of the experiential weight module will lead to a performance decline in average waiting time by 14.89\%.

\begin{table*}[h!]
    \centering
\footnotesize 
    \resizebox{\textwidth}{!}{%
    \begin{tabular}{l@{\hspace{0.7em}}l@{\hspace{0.7em}}l@{\hspace{0.7em}}l@{\hspace{0.7em}}l@{\hspace{0.7em}}l@{\hspace{0.7em}}l@{\hspace{0.7em}}l@{\hspace{0.7em}}l@{\hspace{0.7em}}l}
    \hline
    Model & Metric & Grid $4 \times 4$ & Avenue $4 \times 4$ & Grid $5 \times 5$ & Cologne8 & Ingolstadt21 & Fenglin & Nanshan & Improve\\ \hline
    \multirow{3}{*}{\begin{tabular}[c]{@{}l@{}}DuaLight\end{tabular}} & Delay & 49.32 $\pm$ 0.00 & 756.99 $\pm$ 69.44 & 237.71 $\pm$ 0.00 & 25.35 $\pm$ 0.00 & 182.67 $\pm$ 9.34 & 260.87 $\pm$ 0.00 & 429.49 $\pm$ 0.00& \\
     & Trip Time & 161.04 $\pm$ 0.00 & 396.65 $\pm$ 0.00 & 221.83 $\pm$ 0.00 & 89.74 $\pm$ 0.00 & 317.97 $\pm$ 0.00 & 313.22 $\pm$ 4.88 & 609.89 $\pm$ 0.00& \\
     & Wait & 23.49 $\pm$ 0.00 & 252.43 $\pm$ 3.61 & 83.64 $\pm$ 0.00 & 7.85 $\pm$ 0.17 & 126.91 $\pm$ 0.00 & 157.53 $\pm$ 4.15 & 249.56 $\pm$ 0.00&   \\ \hline
    \multirow{3}{*}{\begin{tabular}[c]{@{}l@{}}w/o \\ Co-Train\end{tabular}} & Delay & 49.27 $\pm$ 0.00 & 804.73 $\pm$ 8.39 & 223.36 $\pm$ 0.00 & 24.86 $\pm$ 0.00 & 221.85 $\pm$ 0.00 & 291.36 $\pm$ 0.00 & 440.46 $\pm$ 0.00 & -4.85\%\\
 & Trip Time & 160.91 $\pm$ 0.00 & 385.11 $\pm$ 0.00 & 214.39 $\pm$ 0.00 & 88.81 $\pm$ 0.00 & 338.07 $\pm$ 0.00 & 346.72 $\pm$ 0.00 & 617.27 $\pm$ 0.00 & -1.55\%\\
 & Wait & 23.48 $\pm$ 0.00 & 226.06 $\pm$ 0.00 & 79.63 $\pm$ 0.00 & 7.56 $\pm$ 0.00 & 148.92 $\pm$ 0.00 & 198.5 $\pm$ 0.00 & 268.71 $\pm$ 0.00 & -4.58\%\\ \hline
 \multirow{3}{*}{\begin{tabular}[c]{@{}l@{}}w/o \\ Exp. Weights \end{tabular}} & Delay & 52.18 $\pm$ 0.07 & 761.52 $\pm$ 19.03 & 207.00 $\pm$ 0.00 & 34.38 $\pm$ 0.00 & 151.07 $\pm$ 0.00 & 266.27 $\pm$ 0.00 & 439.11 $\pm$ 0.00 & -2.30\%\\
     & Trip Time & 163.86 $\pm$ 0.08 & 464.24 $\pm$ 27.07 & 239.66 $\pm$ 0.48 & 98.25 $\pm$ 0.00 & 284.96 $\pm$ 3.30 & 325.73 $\pm$ 0.00 & 618.17 $\pm$ 0.00& -4.47\%\\
     & Wait & 26.21 $\pm$ 0.00 & 304.25 $\pm$ 24.22 & 94.44 $\pm$ 0.00 & 12.81 $\pm$ 0.00 & 96.72 $\pm$ 3.90 & 169.46 $\pm$ 0.00 & 280.07 $\pm$ 0.00&  -14.89\% \\ \hline
    \multirow{3}{*}{\begin{tabular}[c]{@{}l@{}}w/o \\ Intersection-
    \\wise weight\end{tabular}} & Delay & 49.00 $\pm$ 0.51 & 830.84 $\pm$ 37.57 & 222.28 $\pm$ 0.00 & 25.33 $\pm$ 0.08 & 232.69 $\pm$ 16.35 & 274.66 $\pm$ 0.00 & 444.57 $\pm$ 0.00 & -5.53\%\\
     & Trip Time & 160.79 $\pm$ 0.67 & 402.4 $\pm$ 1.04 & 209.26 $\pm$ 0.18 & 89.6 $\pm$ 0.10 & 343.09 $\pm$ 5.99 & 334.91 $\pm$ 0.00 & 615.98 $\pm$ 0.00 & -1.61\%\\
     & Wait & 23.27 $\pm$ 0.00 & 255.02 $\pm$ 0.62 & 78.58 $\pm$ 0.56 & 7.76 $\pm$ 0.21 & 158.14 $\pm$ 5.23 & 183.06 $\pm$ 0.00 & 282.6 $\pm$ 0.00 & -6.71\%\\ \hline
     \multirow{3}{*}{\begin{tabular}[c]{@{}l@{}}w/o \\ Feature-\\wise weight \end{tabular}} & Delay & 49.49 $\pm$ 0.47 & 929.79 $\pm$ 18.56 & 223.53 $\pm$ 0.00 & 25.41 $\pm$ 0.59 & 164.28 $\pm$ 0.57 & 261.33 $\pm$ 5.12 & 444.14 $\pm$ 0.00 & -1.57\%\\
     & Trip Time & 161.02 $\pm$ 0.00 & 471.42 $\pm$ 13.44 & 229.2 $\pm$ 0.00 & 89.45 $\pm$ 0.57 & 289.72 $\pm$ 0.01 & 328.08 $\pm$ 2.85 & 623.51 $\pm$ 0.00 & -2.85\%\\
     & Wait & 23.79 $\pm$ 0.19 & 305.54 $\pm$ 14.57 & 88.8 $\pm$ 0.00 & 7.84 $\pm$ 0.22 & 101.45 $\pm$ 0.14 & 177.98 $\pm$ 0.65 & 277.78 $\pm$ 0.00 & -4.66\%\\ \hline
    \end{tabular}
    }
    \caption{The detailed results of ablation experiments} %\ziyue{also add the full model DualLight}}
    \label{app:table: Ablation Experiment}
\end{table*}

\textbf{Effect of Co-Train}:
The Co-Train module introduces generic information from other scenarios into the \textbf{fe}, \textbf{GAT}, and \textbf{MLP} modules during training. In simple scenarios like \textit{Grid $4 \times 4$}, \textit{Avenue $4 \times 4$}, and \textit{Grid $5 \times 5$}, the model can perform well even without the Co-Train module. Adding it in these cases may introduce redundant information, affecting learning and lowering performance.
However, in complex scenarios like \textit{Nanshan}, \textit{Fenglin}, and \textit{Ingolstadt21}, with irregular road networks and complicated traffic patterns, using the Co-Train module helps the model learn common underlying dynamic information better, resulting in improved performance.

% Regarding the specific effects of each module, using the Co-Train module will introduce generic information from other scenarios into the \textbf{fe}, \textbf{GAT}, and \textbf{MLP} modules during training. For \textit{Grid $4 \times 4$}, \textit{Avenue $4 \times 4$}, and \textit{Grid $5 \times 5$}, the topology of the road network is relatively simple, and traffic patterns are easy to learn. Hence, even without using the Co-Train module, the model can still perform well in these scenarios. Using the Co-Train module in simple scenarios might introduce redundant information from other scenarios, which could interfere with model learning and lead to lower performance. However, for complex scenarios with irregular road network topology and more complicated traffic patterns, such as \textit{Nanshan}, \textit{Fenglin}, and \textit{Ingolstadt21}, introducing information from other scenarios can help the model learn common underlying dynamic information better, which leads to better performance.

\textbf{Effect of Experiential Weight}: 
The Intersection-wise and Feature-wise modules summarize information for each scenario and learn weights for neighbors and individual features. Disabling the weight module hampers scenario-specific learning and reduces performance.
In simple scenarios like \textit{Grid $4 \times 4$} and \textit{Grid $5 \times 5$}, where the road structure is fixed and traffic is even, neighbor information is less important. Disabling the Intersection-wise module improves performance as individual features are sufficient.
However, in complex scenarios like \textit{Ingolstadt21} with intricate road networks and traffic patterns, considering neighbors is crucial for macro-decisions. Disabling the Feature-wise module reduces performance as the GAT module already incorporates individual information effectively.

% \textbf{Effect of Number of Neighbors:} \ziyue{Reviewer 3: Key ablation studies are also missing. It is not clear how different numbers of neighbors affect the model’s performance}

\end{appendices}

%%%%%%%%%%%%%%%%%%%%%%%%%%%%%%%%%%%%%%%%%%%%%%%%%%%%%%%%%%%%%%%%%%%%%%%%

\end{document}